\documentclass[10pt]{article}
\usepackage{fullpage}
\usepackage{graphicx}
\usepackage{cite}
\usepackage{multirow}
\usepackage{verbatim}
\usepackage{amsmath}
\usepackage{amssymb}
\usepackage{color}
\usepackage{stackrel}
\usepackage{upgreek}
\usepackage{csquotes}
\usepackage{rotating}
\usepackage{dcolumn}
\usepackage{cite}
\usepackage[colorlinks=true, allcolors=blue]{hyperref}

\newcolumntype{d}[1]{D{.}{\cdot}{#1} }

\def\red{\textcolor{red}}

\definecolor{lightgray}{gray}{0.8}
\definecolor{llightgray}{gray}{0.95}

\def\##1{{\underline #1}}
\def\=#1{\underline{\underline{#1}}}
\def\+#1{\underline{\bf #1}}
\def\*#1{\breve{\bf #1}}

\newcommand{\SImum}{\ensuremath{\upmu}\textrm{m}}

\def\.{\mbox{ \tiny{$^\bullet$} }}

\def\le{\left(}
\def\ri{\right)}
\def\les{\left[}
\def\ris{\right]}
\def\lec{\left\{}
\def\ric{\right\}}

\def\r#1{(\ref{#1})}

\def\eps{\varepsilon}
\def\epso{\eps_{\scriptscriptstyle 0}}

\def\lambdao{\lambda_{\scriptscriptstyle 0}}

\def\muo{\mu_{\scriptscriptstyle 0}}

\def\Eo{E_{\scriptscriptstyle 0}}
\def\lambdaomin{\lambda_{\scriptscriptstyle 0,min}}
\def\lambdaomax{\lambda_{\scriptscriptstyle 0,max}}

\def\00{^{(0,0)}}

\def\n{^{(n)}}

\def\La{L_{\rm a}}	
\def\Ld{L_{\rm d}}	
\def\Lg{L_{\rm g}}	
\def\Lm{L_{\rm m}}
\def\Ls{L_{\rm s}}
	
\def\Lt{L_{\rm t}}

\def\Lw{L_{\rm w}}
\def\Lx{L_{\rm x}}

\def\Pin{$p$-$i$-$n$}

\def\epsd{\eps_{\rm d}}
\def\epsm{\eps_{\rm m}}
\def\epsg{\eps_{\rm g}}

\def\sfE{{\sf E}}
\def\eg{\sfE_{\rm g}}
\def\ego{\sfE_{\rm g,min}}

\def\egmax{\sfE_{\rm g,max}}

\def\Jsc{J_{SC}}

\def\Voc{V_{oc}}

\def\LCdS{L_{\rm CdS}}
\def\sigman{\sigma_n}
\def\sigmap{\sigma_p}
\def\Pmax{P_{max}}
\def\ni{n_{i}}

\def\Voc{V_{OC}}
\def\ni{n_{\rm i}}
\def\Nc{N_{\rm c}}
\def\Nv{N_{\rm v}}
\def\Nf{N_{\rm f}}

\def\vth{v_{\rm th}}

\def\lambdao{\lambda_{\scriptscriptstyle 0}}

\def\lambdaomin{\lambda_{\scriptscriptstyle 0, {\rm min}}}
\def\lambdaomax{\lambda_{\scriptscriptstyle 0,{\rm max}}}    
\def\La{L_{\rm a}}	
\def\Ld{L_{\rm d}}	
\def\Lg{L_{\rm g}}	
\def\Lm{L_{\rm m}}
\def\Ls{L_{\rm s}}
	
\def\LCdS{L_{\rm CdS}}

\def\Lt{L_{\rm t}}

\def\Lw{L_{\rm w}}
\def\Lx{L_{\rm x}}

\def\eps{\varepsilon}
\def\epso{\eps_{\scriptscriptstyle 0}}

\def\epsd{\eps_{\rm d}}
\def\epsm{\eps_{\rm m}}
\def\epsg{\eps_{\rm g}}
\def\epsdc{\varepsilon_{\rm dc}}

\def\sfE{{\sf E}}
\def\Eo{E_{\scriptscriptstyle 0}}
\def\Ei{\sfE_{\rm i}}
\def\Ec{\sfE_{\rm c}}
\def\Ev{\sfE_{\rm v}}
\def\eg{\sfE_{\rm g}}
\def\ego{\sfE_{\rm g,min}}
\def\egmax{\sfE_{\rm g,max}}
\def\EFn{\sfE_{\rm F_{n}}}
\def\EFp{\sfE_{\rm F_{p}}}

\def\Jsc{J_{\rm sc}}

\def\Jn{J_{\rm n}}
\def\Jp{J_{\rm p}}

\def\Jdev{J_{\rm dev}}

\def\kB{k_{\rm B}}
\def\muo{\mu_{\scriptscriptstyle 0}}
\def\mun{\mu_{\rm n}}
\def\mup{\mu_{\rm p}}
\def\n0{n_{\scriptscriptstyle 0}}
\def\p0{p_{\scriptscriptstyle 0}}
\def\ni{n_{\rm i}}
\def\Nc{N_{\rm c}}
\def\Nv{N_{\rm v}}
\def\Nf{N_{\rm f}}
\def\ND{N_{\rm D}}

\def\RB{R_{\rm B}}

\def\Pmax{P_{\rm max}}
\def\Pin{P_{\rm in}}

\def\qe{q_{\rm e}}

\def\sigman{\sigma_{\rm n}}
\def\sigmap{\sigma_{\rm p}}

\def\Voc{V_{\rm oc}}

\def\Vext{V_{\rm ext}}

\def\Rnpz{R(n,p;z)}
\def\Rnpzrad{R_{\rm rad}(n,p;z)}
\def\RnpzSRH{R_{\rm SRH}(n,p;z)}
\def\RB{R_{\rm B}}

\def\taun{\tau_{\rm n}}
\def\taup{\tau_{\rm p}}

\def\Al2O3{\rm Al_{2}O_{3}}
\def\LAl2O3{L_{Al_{2}O_{3}}}
\def\LZnO{L_{\rm ZnO}}

\def\MgF2{\rm MgF_2}
\def\LMgF2{L_{\rm \MgF2}}

\begin{document}

\begin{center}

\LARGE{ {\bf  Towards highly efficient thin-film solar cells with a graded-bandgap 
CZTSSe layer
}}
\end{center}
\begin{center}
\vspace{10mm} 

Faiz Ahmad,$^1$
 Akhlesh  Lakhtakia,$^{1,2,\ast}$ Tom H. Anderson$^3$ and Peter B. Monk$^3$\\
  \vspace{3mm}
 $^1${NanoMM--Nanoengineered Metamaterials Group, Department of Engineering Science and Mechanics,
Pennsylvania State University, University Park, PA 16802-6812, USA}\\
 \vspace{3mm}
  $^2${Sektion for Konstruktion og Produktudvikling,  Institut for Mekanisk Teknologi, Danmarks Tekniske Universitet, DK-2800 Kongens Lyngby, Danmark}\\
 \vspace{3mm}
$^3${Department of Mathematical Sciences, University of Delaware,
Newark, DE 19716, USA}\\
 \vspace{3mm}
$^\ast${Corresponding author. E--mail: akhlesh@psu.edu}\\

\normalsize

\end{center}

\begin{center}
\vspace{15mm} {\bf Abstract}
\end{center}

 A coupled optoelectronic model was implemented along with the differential evolution algorithm
to assess the efficacy of grading the bandgap of the
Cu$_2$ZnSn(S$_{\xi}$Se$_{1-\xi}$)$_4$ (CZTSSe) layer
for enhancing the power conversion efficiency  of thin-film  CZTSSe solar cells. Both linearly
and sinusoidally graded bandgaps were examined, with the molybdenum backreflector
in the solar cell being either planar or periodically corrugated. Whereas an
optimally graded bandgap can dramatically enhance the efficiency, the effect
of periodically corrugating the backreflector is modest at best.
An efficiency of $21.74$\%  is predicted
with sinusoidal  grading of a $870$-nm-thick CZTSSe layer, 
in comparison to $12.6$\% efficiency achieved experimentally with a $2200$-nm-thick 
homogeneous CZTSSe layer. 
High electron-hole-pair generation rates in the narrow-bandgap 
regions and a high open-circuit voltage due to a wider bandgap close to the front and rear
faces of the CZTSSe layer are responsible for the high enhancement of efficiency.
\vspace{5mm}

\vspace{5mm}

\vspace{10mm}

\section{Introduction}

As the worldwide demand for eco-responsible sources of cheap energy continues to increase for the betterment of an ever-increasing fraction of the human population~\cite{Singh1}, the cost of traditional crystalline-silicon solar cells continues to drop~\cite{Dudley}, as predicted earlier this decade~\cite{Singh2}. While this is a laudable development, small-scale photovoltaic generation of energy must become ubiquitous for human progress to become truly unconstrained by energy economics. 
Thin-film solar cells are necessary for that to happen.

Currently, thin-film solar cells containing absorber layers made
of either CIGS or CdTe are commercially dominant, even over their amorphous-silicon counterparts~\cite{Lee2017}. However, 
there is a strong concern about the planetwide availability of indium (In) and tellurium (Te), both needed for CIGS and CdTe solar cells~\cite{Candelise2012}. 
Furthermore, both In and cadmium (Cd) are toxic, leading to environmental concerns about their impact
following disposal after use. 

Thin-film solar cells must be made from materials that are abundant on our planet and that can extracted, 
processed, and discarded with low environmental cost.
Cu$_2$ZnSn(S$_{\xi}$Se$_{1-\xi}$)$_4$ (commonly referred as CZTSSe) is a $p$-type semiconductor
than can be used in place of   CIGS in a solar cell. 
CZTSSe comprises nontoxic and abundant materials~\cite{Adachi_book2015}. But
the record for the power conversion efficiency $\eta$ of CZTSSe solar cells is only 12.6$\%$~\cite{Wang2013, Wong_survey},
which is substantially lower than the $22.6$\% record efficiency of CIGS solar cells \cite{Jackson2016, Green2018}. 

A low open-circuit voltage $\Voc$ is the
key limitation to high efficiency for CZTSSe solar cells
 \cite{Gershon2014, Ana-Kanevce2015, YLee2015}. This is due to
\begin{itemize}
	\item[(i)] more bandtail states
	in CZTSSe~\cite{Gokmen2013a, Frisk2016};
	\item[(ii)] the \red{high} electron-hole recombination rate inside  the  CZTSSe layer
	because of the short lifetime 
	of  minority carriers (electrons)~\cite{Repins2013};
	and 
	\item[(iii)] the higher electron-hole recombination rate at the CdS/CZTSSe interface~\cite{Gunawan2010}, an ultrathin CdS layer being employed as an $n$-type semiconductor in the solar cell. 
\end{itemize}
The low  lifetime of minority carriers  shortens their  diffusion length, thereby limiting
the collection of minority carriers deep in the CZTSSe absorber layer~\cite{Repins2013, Gokmen2013}. 
For example, the  diffusion length of electrons is less than 1~\SImum~
when the bandgap $\eg$ of CZTSSe is $1.15$~eV (for $\xi\approx 0.41$), which means that a solar cell with a CZTSSe layer
of thickness $\Ls>1$~\SImum~\cite{Gokmen2013} will have a high series resistance~\cite{Gershon2014, Mitzi2011} that will have a deleterious effect on $\eta$. 
Reduction of  $\Ls$ is therefore desirable, all the more so because it will reduce material usage and enhance manufacturing throughput concomitantly. But, a smaller $\Ls$ will reduce the absorption of incident photons.
The common  techniques for tackling this problem in thin-film solar cells are light trapping using nanostructures in front of the illuminated face of the solar cell \cite{Gloeckler-Sites2005,Schmid2017,Schmid2015}, nanostructured 
backreflectors~\cite{Goffard2017}, and back-surface passivation~\cite{Vermang2014}; however, let us note here that enhanced light trapping does not necessarily
translate into higher efficiency~\cite{Ahmad-SPIE2018,Faiz-CIGS-AO}.

The issue of low   $\Voc$, and therefore low $\eta$,  of the CZTSSe solar cell can be tackled by grading the bandgap $\eg$ of the CZTSSe absorber layer in the thickness direction~\cite{Woo2013, Yang2016, Hwang2017,Mohammadnejad2017,Simya2016,Hironiwa2014, Ferhati2018,Chadel2017}.
Since $\eg$ is a function of $\xi\in[0,1]$, the parameter which quantifies the proportion
of sulfur (S) relative to that of selenium (Se) in CZTSSe~\cite{Adachi_book2015, Bag2012, Nakane2016},
the bandgap  can  be graded in the thickness direction by changing $\xi$  dynamically during fabrication ~\cite{Woo2013}. Indeed, bandgap grading of the CZTSSe absorber 
layer has been experimentally demonstrated ~\cite{Woo2013, Yang2016, Hwang2017} to enhance both $\Voc$ and  $\eta$ of  CZTSSe solar cells, but we note that the maximum efficiency reported in Refs.~\citenum{Woo2013,Yang2016,Hwang2017} is $12.3\%$.

The experimental demonstration of increased efficiency due to bandgap grading is supported by theoretical studies. An empirical model recently suggested that a linearly graded $1150$-nm-thick absorber layer can deliver $16.9\%$ efficiency. Several simulations performed with SCAPS software~\cite{Burgelman} have predicted efficiencies between $12.4\%$
and $19.7\%$ with absorber layers between $1000$ and $3500$-nm thickness and the 
bandgap grading being linear \cite{Mohammadnejad2017}, piecewise linear \cite{Hironiwa2014}, parabolic~\cite{Mohammadnejad2017,Chadel2017}, or exponential~\cite{Mohammadnejad2017,Simya2016}. However, the SCAPS software is optically elementary in that it relies on the Beer--Lambert law~\cite{Fonash} rather than on the correct solution of an optical boundary-value problem; a rigorous optoelectronic model is needed to examine bandgap grading for CZTSSe solar cells.

\begin{figure}[!t]
	\centering	
\includegraphics[width=0.8\columnwidth]{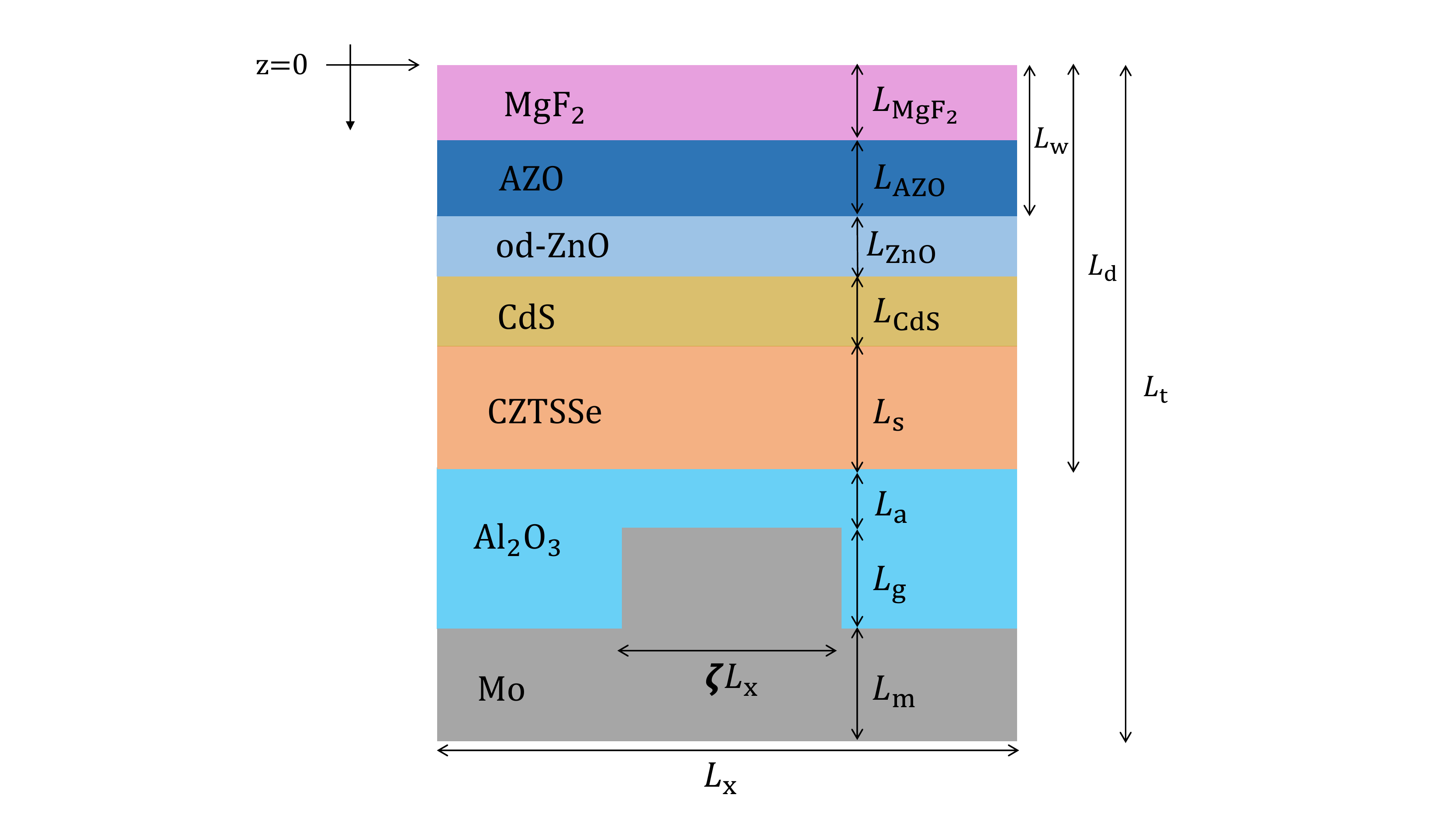} 
	\caption{Schematic of the reference unit cell of the CZTSSe solar cell with a 1D periodically corrugated metallic backreflector.} \label{figure1} 
\end{figure}

A coupled optoelectronic model has recently been devised for CIGS solar cells ~\cite{Faiz-CIGS-AO,Anderson-JCP}. This model was adapted for \red{CZTSSe} solar cells and used
with the differential evolution algorithm (DEA) to maximize $\eta$
for  linear and sinusoidal grading of 
the bandgap of the CZTSSe layer along the 
thickness direction (parallel to the $z$ axis of a Cartesian coordinate system)
in the thin-film solar cell depicted in Fig.~\ref{figure1}. In the optical part of this model,
the rigorous coupled-wave approach (RCWA)~\cite{GG,ESW2013}
is  used to determine
the electron-hole-pair generation rate in the semiconductor
region of the solar cell~\cite{Faiz-CIGS-AO}, assuming normal illumination
 by unpolarized polychromatic light endowed with the AM1.5G solar spectrum~\cite{SSAM15G}. Then, in the electrical part
of the model, the electron-hole-pair generation rate appears as a forcing function in
the one-dimensional (1D) drift-diffusion equations~\cite{Fonash, Jenny_Book} applied
to the semiconductor region. These equations
 are solved using a hybridizable discontinuous Galerkin (HDG) scheme~\cite{Lehrenfeld, CockburnHDG,FuQiuHDG,Brinkman}
to determine the  current density $\Jdev$ and the electrical power density
$P$ as functions of the bias voltage $\Vext$. In turn, the $\Jdev$-$\Vext$ and
the $P$-$\Vext$ curves yield the short-circuit current density $\Jsc$ along with $\Voc$
 and $\eta$.

As shown in Fig.~\ref{figure1}, we took the CZTSSe solar cell to comprise an antireflection
coating of magnesium fluoride ($\MgF2$), followed by
an aluminum-doped zinc oxide (AZO) layer  
as the front contact, a buffer layer of   oxygen-deficient 
zinc-oxide (od-ZnO), the ultrathin CdS layer, and the CZTSSe layer. The od-ZnO, CdS, and CZTSSe layers
constitute the semiconductor region of the solar cell.
Whereas the actual bandgap of CZTSSe was used for the optical part in our calculations, the bandgap was
depressed in the electrical part in order to account
for bandtail defects~\cite{Gokmen2013a, Frisk2016}.
The bandgap-dependent (i.e., $\xi$-dependent) defect density  and electron affinity were used in the electrical 
calculations. The nonlinear Shockley--Read--Hall (SRH) and radiative processes
for electron-hole
recombination were also incorporated~\cite{Fonash, Jenny_Book}.
The Mo backreflector was assumed to be periodically corrugated along a fixed axis (designated
as the $x$ axis) normal to the $z$ axis.
A  thin layer of aluminum oxide ($\Al2O3$) was inserted 
between the CZTSSe layer and the Mo backreflector, as has been experimentally
shown to prevent
the formation of a Mo(S$_\xi$Se$_{1-\xi}$)$_2$ layer that enhances the back-contact 
electron-hole recombination rate and depresses $\eta$~\cite{FLiu2017}. The efficiency
 $\eta$ was maximized  for  (a)
homogeneous, (b)  linearly graded, as well as (c)  sinusoidally graded CZTSSe layers using
the differential evolution algorithm (DEA)~\cite{DEA}.
The role of traps at the  
CdS/CZTSSe interface was assessed by incorporating 
a surface-defect layer
\cite{Songetal2004} with higher defect density.

This paper is organized  as follows.  The optical description
of the  solar cell of Fig.~\ref{figure1}
 is presented in Sec.~\ref{sec:optical-model} along with the approach taken for 
optical calculations. The electrical description of
the solar cell is discussed in Sec~\ref{sec:EDSC}. Optimization for maximum efficiency is briefly discussed in Sec.~\ref{sec:JVEF}.
Section~\ref{sec:Ref-conventional-cell} compares the efficiency of
the   conventional solar cell with a 2200-nm-thick homogeneous CZTSSe layer~\cite{Wang2013} with that   predicted by the coupled optoelectronic model. 
The effects of the $\Al2O3$ layer and
 the CdS/CZTSSe interface recombination rate
on the solar-cell performance are discussed
in Sec.~\ref{sec:Al203_passivation} and Sec.~\ref{sec:SDL}, respectively.
Section~\ref{sec:opto_elechomo_FBR} provides
the optimal configurations of solar cells with a homogeneous  CZTSSe
layer and a planar backreflector,
Sec.~\ref{sec:opto_elechomo_PCBR}  for  solar cells with a homogeneous  CZTSSe
layer and a periodically corrugated backreflector, and
Sec.~\ref{sec:Linearly_nonhomo}  for solar cells
with a linearly graded CZTSSe layer and either a planar or a periodically corrugated backreflector, while
optimal configurations of solar cells with a sinusoidally graded CZTSSe layer and either
a planar or a periodically
corrugated backreflector are presented in Sec.~\ref{sec:optoelec_nonhomo_PCBR}.
Concluding remarks are provided in Sec.~\ref{sec:conc}.

\section{Optoelectronic Modeling and Optimization}	\label{sec:optoelec-model}

\subsection{Optical theory in brief}\label{sec:optical-model}
The CZTSSe solar cell   occupies the region 
${\cal X}:\left\{(x,y,z) \vert -\infty\right.$
$\left. <x<\infty, -\infty<y<\infty, 0<z<\Lt\right\}$, the
 half spaces $z<0$ and $z>\Lt$ being occupied by air.
The reference unit cell of this structure, shown in Figure~\ref{figure1}, occupies the region 
${\cal R}:\left\{(x,y,z) \vert -\Lx/2<x<\Lx/2,  -\infty<y<\infty, 0<z<\Lt\right\}$.
The  region $0<z<\Lw=210$~nm consists of a 110-nm-thick antireflection coating~\cite{Rajan} made
of $\MgF2$ layer~\cite{mgf2}   and a 100-nm-thick AZO layer~\cite{AZO} 
as the front contact. The region $\Lw<z<\Lw+\LZnO$ is a 100-nm-thick buffer layer of oxygen-deficient 
zinc oxide (od-ZnO)~\cite{ZnO}. 
Oxygen deficiency during the deposition of ZnO makes it an $n$-type semiconductor \cite{Wellings}. The region $\Lw+\LZnO<z<\Lw+\LZnO+\LCdS$ is a 50-nm-thick layer of $n$-type 
CdS~\cite{treharne} that forms a junction with the $p$-type CZTSSe layer of thickness $\Ls\in[100,2200]$~nm 
and a bandgap $\eg(z)$ that can vary with $z$. 
The  region $\Ld+\La+\Lg<z<\Lt$ is occupied by Mo~\cite{Mo} of permittivity $\epsm(\lambdao)$, 
where $\lambdao$ is the free-space wavelength.
The thickness $\Lm=\Lt-\left(\Ld+\La+\Lg\right)=500$~nm was chosen to be significantly larger than the electromagnetic
penetration depth \cite{Iskander} of Mo across the visible spectrum. A thin $\Al2O3$~\cite{Al2O3} layer of thickness $\La=20$~nm and permittivity  $\epsd(\lambdao)$ exists between the Mo backreflector and 
the CZTSSe absorber layer~\cite{FLiu2017}.

\begin{figure}[!t]
	\centering	
	\includegraphics[width=0.9\columnwidth]{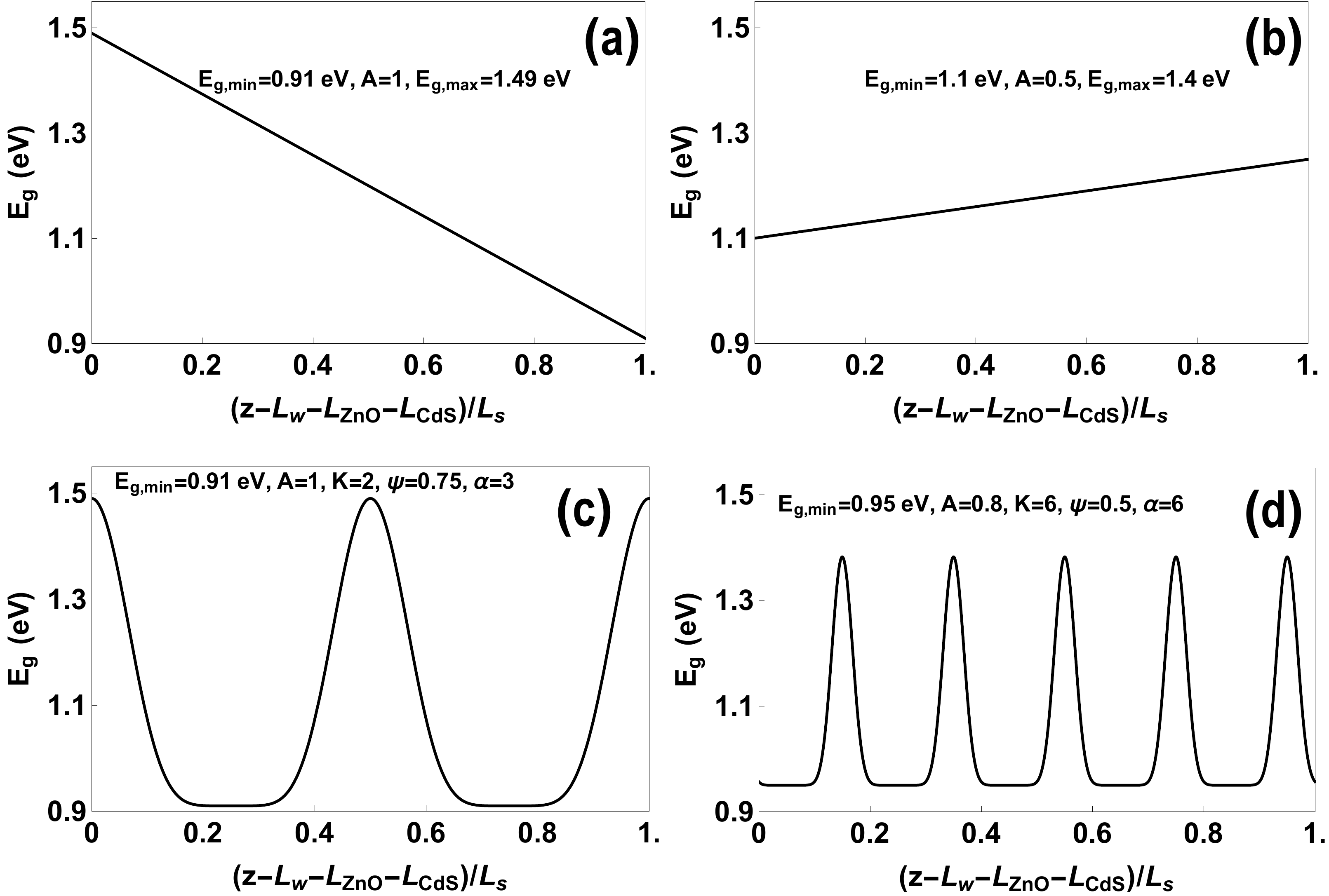} 
	\caption{Representative profiles of (a,b) linearly 
	and (c,d) sinusoidally graded bandgaps in a CZTSSe   layer 
		with different \red{sets} of parameters.} 
		\label{figure2}  
\end{figure}

The region $\Ld+\La<z<\Ld+\La+\Lg$ has a rectangular 
metallic grating with period $\Lx$ along the $x$ axis. In this region,
the permittivity  is given by
\begin{eqnarray}
	\nonumber
	&&
	\epsg(x, z,\lambdao) =\left\{\begin{array}{ll}\epsm(\lambdao) \,,&\qquad \vert{x}\vert<\zeta \Lx/2\,,
		\\[4pt] \epsd(\lambdao) \,,& 
		\qquad \vert{x}\vert\geq\zeta \Lx/2\,,
	\end{array}\right.
	\\
	&&		
	\quad z\in(\Ld+\La,\Ld+\La+\Lg)\,,
	\label{Eqn:Grating}
\end{eqnarray}
with $\zeta\in(0,1)$ as the duty cycle. The grating is absent for $\zeta\in\left\{0,1\right\}$.

The linearly nonhomogeneous bandgap can be either backward graded or
forward graded. For backward grading,
\begin{eqnarray}
	\nonumber
	&&
	\eg(z)=\egmax 
	\\&&-A\left(\egmax-\ego\right)\frac{z-\left(\Lw+\LZnO+\LCdS\right)}{\Ls}\, , 
	\nonumber
	\\ &&\qquad z\in\left[\Lw+\LZnO+\LCdS, \Ld\right]\, ,
	\label{Eqn:Linear-bandgap}
\end{eqnarray}
where $A$ is an amplitude, $\ego$ is the minimum
bandgap, and $\egmax$ is the maximum bandgap;
 $A = 0$ represents a homogeneous CZTSSe layer. 
For forward     grading,
\begin{eqnarray} 
	\nonumber
	&&\eg(z)=\ego 
	\\ &&+A\left(\egmax-\ego\right)\frac{z-\left(\Lw+\LZnO+\LCdS\right)}{\Ls}\, ,
	\nonumber
	\\
	&&
	\qquad z\in\left[\Lw+ \LZnO+\LCdS, \Ld\right]\, .
	\label{Eqn:Linear-bandgap1}
\end{eqnarray} 
Two representative bandgap profiles are shown in  Figs.~\ref{figure2}(a,b).

For the sinusoidally graded bandgap,
\begin{align} 
	\nonumber
	&\eg(z)=\ego +A\le 1.49-\ego\ri \, 
	\\
	\nonumber
	&\times
	\lec \frac{1}{2}\, \les \sin\le 2\pi K \frac{z-\left(\Lw+\LZnO+\LCdS\right)}{\Ls}-2\pi \psi\ri +1\ris\, \ric^{\alpha} \, , 
	\\[5pt]
	&\qquad\qquad  z\in\left[\Lw+\LZnO+\LCdS, \Ld\right]\, ,
	\label{Eqn:Sin-bandgap}
\end{align}
where the integer $K$ is the number of periods in the CZTSSe layer, $\psi\in [0, 1]$ describes a relative phase shift, and $\alpha> 0 $ is a
shaping parameter. 
Two representative profiles are provided in Figs.~\ref{figure2}(c,d). As thin-film solar cells are fabricated using vapor-deposition techniques~\cite{RJMP}, graded bandgap profiles could be physically realized by adjusting the sulfur-to-selenium ratio in the precursor and thus
varying the composition parameter  $\xi \in[0,1]$ 
during the deposition process~\cite{Wei2011,Woo2013, Yang2016}.
Optical spectra of the relative permittivities of  all materials used in the solar cell
are provided in Appendix \ref{Appendix A}.

The RCWA ~\cite{GG, ESW2013} was used 
for monochromatic calculations. The electric field phasor ${\#E}(x,z,\lambdao)$
and the magnetic field phasor ${\#H}(x,z,\lambdao)$, created
everywhere inside the solar cell due to illumination by an unpolarized  plane wave normally incident on the plane $z=0$, were calculated with  $\Eo=4\sqrt{15 \pi}$~V~m$^{-1}$ being the amplitude
of the incident electric field.
The region $\cal R$ was partitioned into a 
sufficiently large number of slices along the $z$ axis, in order to
implement the RCWA. Although each slice was   homogeneous along the $z$ axis,
it could be periodically nonhomogeneous along the $x$ axis. The slice thickness was chosen by trial and error
such that the useful solar absorptance~\cite{Ahmad2018}
converged with a preset tolerance of $\pm1\%$.  The usual boundary conditions
on the continuity of the tangential components
of the electric and magnetic field phasors
were enforced on the plane $z=0$ to match the internal field phasors to the
sum of the incident and reflected field phasors. The same was done 
 to match the internal field phasors to the
transmitted field phasors on the plane
 $z=\Lt$. Detailed descriptions of the RCWA for solar cells are available~\cite{Ahmad2018, ESW2013, Anderson2018}. 

Suppose that every  photon absorbed in the
semiconductor region $\Lw<z<\Ld$ excites an electron-hole pair. Then,
the $x$-averaged electron-hole-pair generation rate   is~\cite{Faiz-CIGS-AO}        
\begin{eqnarray} 
\nonumber    
&& G(x)=\frac{\sqrt{\muo/\epso}}{\hbar\Eo^2 \Lx } \int_{-\Lx/2}^{\Lx/2}\Bigg[ 
\int_{\lambdaomin}^{\lambdaomax} 
{\rm Im}\{\eps(x, z, \lambdao)\}
\\
&&
\qquad \times\left\vert\#E(x, z, \lambdao)\right\vert^2\,
S(\lambdao)  \, d\lambdao\Bigg]
dx\,
\label{G1D-def}
\end{eqnarray}    
for
$z\in\les \Lw, \Ld \ris$,
where $\hbar$ is  the reduced Planck constant, $\epso$ is the free-space permittivity, $\muo$ is the free-space permeability,
$S(\lambdao)$ is the AM1.5G solar spectrum~\cite{SSAM15G},
$\lambdaomin = 300$~nm, and  $\lambdaomax = \left(1240/\ego\right)$~nm with $\ego$ stated in eV.
Averaging about the $x$ axis can be justified for two reasons. First, 
any current generated parallel to the $x$ axis shall be negligibly small
 because the solar cell operates under the influence of a $z$-directed electrostatic field due
 to the
application of   $\Vext$.  Second, for electrostatic analysis  \red{$\Lx\sim$500~nm
  is} very small in comparison to the lateral dimensions of the solar cell.

\subsection{Electrical theory in brief}\label{sec:EDSC}
For electrical modeling, only the  region $\Lw<z<\Ld$  
has to be considered,  because electron-hole pair generation occurs in the CZTSSe, od-ZnO, and CdS layers only.
With a bandgap of 3.3~eV, od-ZnO
 absorbs solar photons with energies corresponding
to $\lambdao\in\les 300,376\ris$~nm. Likewise,   CdS  absorbs solar photons with energies corresponding
to $\lambdao\in\les 300,517\ris$~nm, as its
bandgap is 2.4~eV.  The planes $z=\Lw$ and
$z=\Ld$ were assumed to
be ideal ohmic contacts, as
we are not interested in how the solar cell interacts with an external circuit.

The electron quasi-Fermi level
	\begin{equation}
	\EFn(z)=\Ec(z)+\red{\left(\kB T\right)} \ln \les n(z)/\Nc(z)\ris
	\label{EFn-def}
	\end{equation}
and the hole quasi-Fermi level
	\begin{equation}
	\EFp(z)=\Ev(z)-\red{\left(\kB T\right)}  \ln \les p(z)/\Nv(z)\ris
	\label{EFp-def}
	\end{equation}
depend on  
$\Nc(z)$ as the density of states in the
conduction band, $\Nv(z)$ as the density of states in the valence band,
$\Ec(z)={\sf E}_0-\les\red{\qe}\phi(z)+\chi(z)\ris$  as the conduction band-edge energy,
$ \Ev(z)=\Ec(z)-\eg(z)$ as the valence band-edge energy,  $\phi(z)$ as 
the dc electric potential,   $\chi(z)$ as the
bandgap-dependent electron affinity, $\kB$ as the Boltzmann constant, and $T$ as
the absolute temperature.
The reference energy level ${\sf E}_0$ is arbitrary.

The gradients of the quasi-Fermi levels drive the 
the electron-current density $\Jn(z)$ and the hole-current density $\Jp(z)$; thus,
\begin{equation}
	\left.\begin{array}{l}
		\Jn(z) = \displaystyle{\red{\mun\, n(z)  \frac{d}{dz}\EFn(z) }}
		\\[6pt]
		\Jp(z) = \displaystyle{\red{\mup\, p(z)  \frac{d}{dz}\EFp(z)}}
	\end{array}\right\}\,,
	\quad
	z\in(\Lw,\Ld)
	\label{Eqn:JnJp},
\end{equation}
where $\qe = 1.6\times$10$^{-19}$~C is the   charge quantum,
$n(z)$ is   the   electron density,
$p(z)$ is   the     hole density, 
$\mun$ is  the electron mobility, and
$\mup$  is  the  hole mobility.
According to the Boltzmann approximation~\cite{Jenny_Book},  
\begin{equation}
\left.\begin{array}{l}
n(z) = \displaystyle{ \ni(z) \red{\exp\lec\les{\EFn (z)- \Ei(z)}\ris/{\kB T}\ric}}
\\[5pt]
p(z) = \displaystyle{\ni(z)\red{\exp\lec-\les{\EFp(z) - \Ei(z)}\ris/{\kB T}\ric}}
\end{array}\right\}\,,
\label{Eqn:ndef-pdef}
\end{equation}
where   
\begin{equation}
\ni (z) = \sqrt{\Nc(z) \,\Nv(z)}\, \exp\les \red{-} {\eg(z)}/{2\kB T}\ris
\end{equation}
is the intrinsic charge-carrier density 
 and
\begin{equation}
\Ei(z)=(1/2) 
\lec\Ec(z)+\Ev(z) -\red{\left(\kB T\right)}  \ln\les\Nc(z)/\Nv(z)\ris\ric
\end{equation}
is the intrinsic energy. Both
$\Nc(z)$ and $\Nv(z)$  are functions of $z$ because they depend
on the bandgap but we took them to be independent of $z$, following Hironiwa \textit{et al.}~\cite{Hironiwa2014}, 
because  bandgap-dependent values are unavailable for CZTSSe.

The 1D drift-diffusion model  comprises the  three differential equations \cite[Sec.~4.6]{Jenny_Book}
\begin{equation}
\left.\begin{array}{l}
\displaystyle{\frac{d}{dz}\Jn(z)}= -\qe\les G(z) - \Rnpz\ris \\[8pt]
\displaystyle{\frac{d}{dz}\Jp(z)} = \qe\les G(z) - \Rnpz\ris \\[8pt]
\displaystyle{\epso\frac{d}{dz}\les\epsdc(z)  \frac{d}{dz}\phi(z)\ris} =
- \qe\les\Nf(z) + \red{\ND(z) +} p(z) - n(z)\ris
\end{array}
\right\}\,,\quad
	z\in(\Lw,\Ld)\,,
	\label{Eqs-1DD}
\end{equation}
under steady-state conditions, 
with
$\Rnpz$  as the electron-hole-pair recombination rate,
$\Nf(z)$ as  the defect density or the trap density, \red{$\ND(z)$ is the doping density
which is positive for donors and negative for acceptors,}
and 
$\epsdc (z)$   as the dc relative permittivity. Both  $\Nf$ and $\epsdc$ depend
on  $\xi$ and, therefore, on $\eg$.
All three differential equations have to be solved simultaneously for $z\in(\Lw,\Ld)$.

The radiative recombination rate is given by
\begin{equation}
\Rnpzrad = {\RB} \les n(z)p(z) - \ni^2(z)\ris,\label{rad}
\end{equation}
where $\RB$ is the radiative recombination coefficient \cite{Jenny_Book, Fonash}. 
The  SRH recombination rate is given by
\begin{eqnarray}
\nonumber
&&
\RnpzSRH  = 
\\
&&\frac{n(z)p(z) - \ni^2(z)}{\taup(z)\les n(z) + n_1(z)\ris + \taun(z)\les p(z)+p_1(z)\ris}\,,\label{SRH}
\end{eqnarray}
where $n_1(z)$ is the electron density and    $p_1(z)$ is the
hole density  at the trap energy level $\sfE_{\rm T}$; 
the electron lifetime
$\taun(z)= {1}/{\les\sigman \vth \Nf(z)\ris}$
depends on   the electron-capture cross section $\sigman$,
the hole lifetime
$\taup(z)= {1}/{\les\sigmap \vth \Nf(z)\ris}$
depends on   the hole-capture cross section $\sigmap$,
and 
$\vth$ is the mean thermal speed of all charge carriers \cite{Jenny_Book, Fonash}. The total recombination rate then is $\Rnpz =\Rnpzrad+\RnpzSRH$.

Dirichlet boundary conditions on $n(z)$, $p(z)$, and
$\phi(z)$ at the planes $z=\Lw$ and $z=\Ld$ supplement
Eqs.~\r{Eqs-1DD}   \cite{Anderson2018, Anderson-JCP}. These boundary conditions were derived after assuming
the region $\Lw<z<\Ld$ to be charge-free and in local quasi-thermal equilibrium~\cite{Fonash}.
The  bias voltage $\Vext$ was taken to be applied at the plane $z=\Ld$. 

The HDG scheme~\cite{Chen2016, CockburnHDG, FuQiuHDG} was used to solve all three
differential equations.  This scheme works well  for 
 solar cells containing heterojunction interfaces~\cite{Brinkman}.  All 
$z$-dependent variables are discretized in this scheme
using discontinuous finite elements in a space of piecewise polynomials
of a fixed degree. We used the Newton--Raphson method to solve the resulting system
for $n(z)$, $p(z)$, and $\phi(z)$~\cite{Brezzi2002}.  

Table~\ref{CZTSSe-data} provides the values of electrical parameters used for od-ZnO, CdS, and CZTSSe~\cite{Frisk14, Adachi_book2015, Ana-Kanevce2015}.
The effect of bandtail states, which effectively narrow the bandgap, was incorporated~\cite{Gokmen2013a, Frisk2016}
by reducing the bandgap of CZTSSe for electrical calculations. 
Whereas $\eg=0.91+0.58\xi  \in[0.91,1.49]$~eV was used for CZTSSe in the optical part
of the coupled optoelectronic model,
$\eg=0.91+0.44\xi \in[0.91,1.35]$~eV was used in the electrical part~\cite{Frisk2016, Gokmen2013a}.

\begin{table}[!t]
	\caption{Electrical properties of od-ZnO, CdS, and CZTSSe for $\xi\in[0,1]$. \label{CZTSSe-data}}
	\centering
	\begin{tabular}{|p{2.8cm}| p{1.8cm}|p{1.8cm}|p{4.6cm}|} \hline
		\red{Symbol (unit)} &od-ZnO \cite{Frisk14}& CdS \cite{Frisk14}&CZTSSe \cite{Adachi_book2015, Ana-Kanevce2015}\\
		\hline\hline
		$\eg$~(eV) &3.3&2.4 \multirow{6}{*}{}& $0.91 + 0.58\xi$ (optical part)\,  \\  
		\cline{4-4}
		 & &  & $0.91 + 0.44\xi$ (electrical part)$^\dag$\,  \\  \hline

		$\chi$~(eV)&4.4&4.2 & $4.46 - 0.16\xi$ \,  \\ \hline 	
		\red{$\ND$}~(cm$^{-3}$)&1$\times10^{17}$\, (donor)  &5$\times10^{17}$\, (donor) & 1$\times10^{16}$\, (acceptor)\\ \hline
		$\Nc$~(cm$^{-3}$)&$3\times 10^{18}$& $1.3\times 10^{18}$&  $7.8\times 10^{17}$ \\ \hline
		$\Nv$~(cm$^{-3}$)&$1.7\times 10^{19}$ &$9.1\times 10^{19}$ &$4.5\times 10^{18}$ \\ \hline
		$\mun$~(cm$^2$~V$^{-1}$~s$^{-1}$)&$100$ &$72$ &40\\ \hline
		$\mup$~(cm$^2$~V$^{-1}$~s$^{-1}$)&31 &$20$ &12.6\\ \hline	
		$\epsdc$&9&5.4 & $14.9 - 1.2\xi$\, \\ \hline	
		$\Nf$~(cm$^{-3}$)&$10^{16}$ &$10^{12}$ & $(1.35 + 98.65\xi)\times10^{15}$\\ \hline	
		$\sfE_{\rm T}$& midgap&midgap & midgap\\ \hline	
		$\sigman$~(cm$^2$)&$5\times10^{-13}$& $5\times10^{-13}$& $10^{-14}$\\  \hline	
		$\sigmap$~(cm$^2$)& $10^{-15}$& $10^{-15}$& $10^{-14}$\\ \hline
		$\RB$~(\red{cm$^{3}$}~s$^{-1}$)&$10^{-10}$ & $10^{-10}$&$10^{-10}$  \\ \hline
		$\vth$~(cm~s$^{-1}$) & $10^7$ &  $10^7$ &  $10^7$
		\\ \hline\hline
	\end{tabular}\\
	$^\dag$ $\eg$ is artificially reduced in the electrical part so as to account for bandtail states.
\end{table}

\subsection{Optoelectronic optimization} \label{sec:JVEF}	
The total current density $\Jn(z) + \Jp(z)$ equals $\Jdev$
everywhere in the od-ZnO, CdS, and CZTSSe layers, under steady-state conditions. When
the solar cell is connected to an external circuit,
$\Jdev$ is the current density delivered by the former to the latter. The short-circuit current density
$\Jsc$ is the value of $\Jdev$ when $\Vext=0$ and the open-circuit voltage
$\Voc$ is the value of $\Vext$ such that $\Jdev=0$. The power density is defined as
$P=\Jdev\Vext$; the maximum power density
$\Pmax$  obtainable from the solar cell is the highest value of $P$ on the $P$-$\Vext$ curve;
and $\eta = {\Pmax}/{\Pin}$, where $\Pin = 1000$~W~m$^{-2}$   is the integral of  $S(\lambdao)$ over the solar spectrum.  The  fill factor $FF={\Pmax}/\left({\Voc\Jsc}\right)\in[0,1]$ is commonly encountered in the solar-cell literature.

The DEA \cite{DEA} was used to optimize $\eta$ with respect to certain geometric and bandgap parameters, 
using a custom algorithm implemented 
with MATLAB\textsuperscript{\textregistered} version R2017b.  

\section{Numerical results and discussion}\label{sec:OptoElecRes}

\subsection{Conventional CZTSSe solar cell (model validation)}\label{sec:Ref-conventional-cell}
  Our coupled optoelectronic model was validated by comparison
with  experimental results available for
the conventional $\MgF2$/AZO/od-ZnO/CdS/CZTSSe/Mo(S$_\xi$Se$_{1-\xi}$)$_2$/Mo solar cell
containing a 2000-nm-thick homogeneous CZTSSe layer and a planar backreflector~\cite{Wang2013}. 
In this solar cell,
a 200-nm-thick Mo(S$_\xi$Se$_{1-\xi}$)$_2$  layer with   defect density $\Nf=10^{18}$ cm$^{-3}$
is present whereas the $\Al2O3$ layer is absent in relation to Fig.~\ref{figure1},
and we made appropriate modifications for the validation. All other relevant electrical parameters
of Mo(S$_\xi$Se$_{1-\xi}$)$_2$ were taken to be the same as that of CZTSSe,
except that $\eg = 1.57+0.31\xi$~eV~\cite{Beal-Hughes} was used for Mo(S$_\xi$Se$_{1-\xi}$)$_2$
in both the optical and electric parts of the coupled
optoelectronic model.
The relative permittivity of Mo(S$_\xi$Se$_{1-\xi}$)$_2$ in the optical regime is  
provided in Appendix~\ref{Appendix A}.

The values of  $\Jsc$, $\Voc$,   $FF$, and $\eta$ obtained from our coupled
optoelectronic model for
$\xi\in\left\{0,0.38,1\right\}$ are provided in 
Table~\ref{tab--ref-results} along with the corresponding experimental data~\cite{Mitzi2011, Wang2013, Frisk2016}. According to this table,
the model's predictions are in reasonable agreement with the experimental data,
the variances being very likely due to differences between the optical and electrical properties inputted
to the model from those realized in practice. As interface defects are not explicitly considered in our model,   all
the experimentally observed features can be adequately
accounted for by the bulk properties of CZTSSe, which is also
 in accord with the empirical model provided by Gokmen \textit{et al.}~\cite{Gokmen2014}.

In order to further elaborate the role of the Mo(S$_\xi$Se$_{1-\xi}$)$_2$ layer, we lowered its thickness  from $200$~nm to $100$~nm but increased the thickness $\Ls$
of the CZTSSe absorber layer from $2000$~nm to $2100$~nm.  The composition
parameter $\xi$ was taken to be $0.38$ for the Mo(S$_\xi$Se$_{1-\xi}$)$_2$ layer as well as for the 
CZTSSe layer,
but other parameters remained the same as for the model's results stated in Table~\ref{tab--ref-results}. The model-predicted efficiency increased from $11.15\%$ to $11.23\%$, indicating the minor role of  the thickness of the Mo(S$_\xi$Se$_{1-\xi}$)$_2$ layer.
%

\begin {table}[!t]
\caption {\label{tab--ref-results}
	{Comparison of   $\Jsc$, $\Voc$,   $FF$, and $\eta$ predicted by
		the coupled optoelectronic model 
		for a conventional $\MgF2$/AZO/od-ZnO/CdS/CZTSSe/Mo(S$_\xi$Se$_{1-\xi}$)$_2$/Mo solar cell with 
		a homogeneous 2000-nm-thick CZTSSe layer (i.e., $A=0$), a 
		homogeneous 200-nm-thick Mo(S$_\xi$Se$_{1-\xi}$)$_2$ layer,
		and a planar backreflector
		with  experimental counterparts. The $\Al2O3$ layer is absent for these data.}
}
\begin{center}
	\begin{tabular}{|p{0.8cm}|p{1.8cm}|p{1.8cm}|p{.8cm}|p{.8cm}|p{.8cm}|}
		\hline \hline
		$\xi$&     & $\Jsc$ &$\Voc$ & $FF$&$\eta$  \\
		&   &(mA~cm$^{-2}$) &(mV)&(\%)& (\%)\\
		\hline 
		0     & \multirow{6}{*}{}Model &38.31&361& 65& 8.96\\ 
		\cline{2-6}
		&Experiment     & & & & 	\\ 
		&~(Ref.~\citenum{Mitzi2011})  &36.4&412&62&9.33	\\ 
		 \hline
		0.38  & \multirow{5}{*}{}Model  &32.42&509& 69& 11.15\\ 
		\cline{2-6}
		&Experiment      & & & & 	\\ 
		& ~(Ref.~\citenum{Wang2013})  &35.2&513.4& 69.8& 12.6\\ 
		\hline	
		1   & \multirow{5}{*}{}Model &17.86&606&60.7& 6.61 \\
		\cline{2-6}
		&Experiment      & & & & 	\\ 
		& ~(Ref.~\citenum{Frisk2016})  &16.9&637&61.7&6.7	\\ 
		\hline \hline
	\end{tabular}
\end{center}
\end {table}	

\subsection{Effect of $\Al2O3$ layer}\label{sec:Al203_passivation}

The incorporation of an ultrathin $\Al2O3$ layer below the CZTSSe layer prevents
the formation of a Mo(S$_\xi$Se$_{1-\xi}$)$_2$ layer and thereby enhances performance~\cite{FLiu2017}. Removing
the  Mo(S$_\xi$Se$_{1-\xi}$)$_2$ layer and
reverting to the solar cell depicted in Fig.~\ref{figure1}, we optimized the CZTSSe solar cell
with and without a $20$-nm-thick $\Al2O3$ layer between the CZTSSe layer
and a planar Mo backreflector. 

Values of  $\Jsc$, $\Voc$,   $FF$, and $\eta$ obtained from our coupled optoelectronic model 
for $\Ls=2200$~nm 
are presented in Table~\ref{tab--Al2O3-paasivation}. 
The optimal efficiency is $11.76$\% with the $\Al2O3$ layer and $11.37$\% without it.  Thus,
the  $\Al2O3$ layer enhances $\eta$ slightly, and concurrent improvements in both  $\Jsc$ and  $\Voc$  can also be noted in Table~\ref{tab--Al2O3-paasivation}. 
Hence, the 20-nm-thick $\Al2O3$ layer
was incorporated in the solar cell for all of the following results.

\begin {table}[!t]
\caption {\label{tab--Al2O3-paasivation} 
	{Predicted parameters of the optimal 
	$\MgF2$/AZO/od-ZnO/CdS/CZTSSe/$\Al2O3$/Mo solar cell
		with and without the $\Al2O3$  layer when the 2200-nm-thick CZTSSe layer is homogeneous
		($A=0$ and $\ego\in[0.91, 1.49]$~eV for the optical part) and the Mo backreflector is planar ($\Lg=0$).
		}
}
\begin{center}
	\begin{tabular}{|p{0.8cm} |p{0.8cm}|p{1.8cm}|p{0.8cm}| p{0.8cm}|p{0.8cm}|}
		\hline
		\hline
		$\La$  &  $\xi$  & $\Jsc$ &$\Voc$ & $FF$&$\eta$  \\
		(nm) &  &(mA~cm$^{-2}$) &(mV)&(\%)& (\%)\\ 
		\hline
		0   &0.50 &29.51&552& 69.70& 11.37\\ \hline
		20  &0.50  &30.00&557&70.31& 11.76\\ 
		\hline \hline
	\end{tabular}
\end{center}
\end {table}


\subsection{Effect of surface recombination on CdS/CZTSSe interface }\label{sec:SDL}

A 10-nm-thin surface-defect layer  was inserted between the CdS and CZTSSe 
layers to investigate the effect of surface recombination at that interface
on the performance of the solar cell depicted in Fig.~\ref{figure1}. The CZTSSe layer
was taken to be homogeneous with thickness $\Ls\in\les100,2200\ris$~nm and all other parameters
as reported in Table~\ref{CZTSSe-data}.
The  surface defect density was fixed at $10^{12}$~cm$^{-2}$ but the mean thermal speed was varied between $10^2$ cm s$^{-1}$ 
and $10^{7}$ cm s$^{-1}$ in the surface-defect layer, with all other characteristics of this layer  taken
to be the same as of the CZTSSe layer.  

The optimal value of $\xi=0.51$ for $\Ls=100$~nm. On inserting the surface-defect layer,
the efficiency reduced from $7.41$\% to: (i) $7.31$\%   when $\vth=10^2$~cm~s$^{-1}$ in the
surface-defect layer and (ii) $7.22$\% when  
$\vth$ is 10$^{7}$~cm~s$^{-1}$ in the
surface-defect layer. 
The optimal value of $\xi=\red{0.50}$ for $\Ls=2200$~nm. On inserting the surface-defect layer,
the efficiency reduced from $11.15$\% to: (i) $11.07$\%   when $\vth=10^2$~cm~s$^{-1}$ in the
surface-defect layer and (ii) $11.02$\% when  
$\vth$ is 10$^{7}$~cm~s$^{-1}$ in the
surface-defect layer. Similar efficiency reductions were predicted for
intermediate values of $\Ls$.
These efficiency reductions are so small that the surface-defect layer can be ignored with minimal consequences. 
Therefore, we neglected surface recombination on the CdS/CZTSSe interface for all results presented from now onwards. 

\subsection{Optimal solar cell: Homogeneous bandgap \& planar backreflector}\label{sec:opto_elechomo_FBR}

Next, we  optimized a  solar cell in which the CZTSSe layer is homogeneous
($A=0$) and the backreflector is planar ($\Lg=0$), in order to highlight the advantage of the nonhomogeneous CZTSSe layer. 

For a fixed value of $\Ls$,  the parameter space
for  optimizing $\eta$  is:   $\ego\in[0.91, 1.49]$~eV (for the optical part\footnote{Throughout Sec.~\ref{sec:OptoElecRes}, the values
of $\eg$ stated for the
CZTSSe layer pertain to the optical part of the coupled optoelectronic model. Knowing  
$\eg$ for the optical part, one can use Table~\ref{CZTSSe-data} to find $\xi$ and, therefore, $\eg$ for the electrical part
of the coupled optoelectronic model.
}).
With $\Ls=2200$~nm, the maximum efficiency predicted for $A=0$ and $\Lg=0$
is $11.76\%$ when  $\ego=1.20$~eV. The corresponding values of $\Jsc$, $\Voc$, and  
$FF$  are $30.00$~mA~cm$^{-2}$, $557$~mV, and $70.3\%$, respectively. Incidentally,
the  efficiency becomes lower for   $\ego>1.2$~eV because of
\begin{itemize}
\item [(a)] the narrowing of  the portion of the solar spectrum available
for photon absorption~\cite{SQlimit}  due to the blue shift of
$\lambdaomax$, and
\item [(b)] the increased recombination due to increase in $\Nf$ caused 
by the higher value of $\xi$~\cite{Adachi_book2015, Ana-Kanevce2015, Mitzi2011}. 
\end{itemize}
 
Next, we considered $\Ls\in[100,2200]$~nm also as a parameter for maximizing $\eta$. 
The highest efficiency predicted is $11.84\%$, produced by a solar cell with a 1200-nm-thick CZTSSe layer
with an optimal bandgap of $\ego=1.21$ eV. The values of $\Jsc$, $\Voc$, and  $FF$ corresponding to this 
optimal design are $30.13$~mA~cm$^{-2}$, $558$~mV, and $70.3$\%, respectively.

In order to compare the performance of the solar cell with optimal $\Ls$,
values of    $\ego$ (for the optical part), $\Jsc$, $\Voc$, 
$FF$, and $\eta$ predicted by the coupled optoelectronic  model are presented in Table~\ref{tab-homo-FBR} for seven 
representative values of $\Ls$.   The maximum efficiency increases to 11.84\% as $\Ls$ increases to 1200~nm, but  decreases at a very slow rate with further increase of $\Ls$. The efficiency increase with $\Ls$ for $\Ls<1200$~nm is due to the increase in volume available to absorb photons.
The efficiency reduction for $\Ls> 1200$~nm is due to reduced charge-carrier collection arising from
short diffusion length of minority charge carriers in CZTSSe being smaller than 
$\Ls$~\cite{Gokmen2013}.  Notably, the optimal
bandgap of
the CZTSSe layer fluctuates in a small range (i.e., $[1.18,1.21]$~eV), despite a $22$-fold increase of $\Ls$.

\begin {table}[h]
\caption {\label{tab-homo-FBR} 
	Predicted parameters of the optimal CZTSSe solar cell with a specified value of $\Ls\in[100,\red{2200}]$~nm, 
	when the CZTSSe layer is homogeneous ($A=0$) and and the Mo backreflector is planar ($\Lg=0$). The values of $\ego$ provided pertain to the optical part of the model. } 
\begin{center} 
	\begin{tabular}{ |p{1.0cm}| p{1.0cm}|p{1.8cm}|p{1.0cm}|p{1.0cm}|p{1.0cm}|}
		\hline
		\hline
		$\Ls$&  $\ego$ &	$\Jsc$ & $\Voc$&$FF$&$\eta$  \\
		(nm) &   (eV) &(mA~cm$^{-2}$) & (mV)& (\%)& (\%) \\
		\hline
		100 &  1.21  & 19.23 & 513   & 75.2  & 7.41	\\  \hline
		200 &  1.20 & 25.19 & 535   & 72.0  & 9.67\\ \hline
		300 &  1.20 & 27.27 & 546   & 69.6  & 10.38\\ \hline
		400 &  1.20 & 28.07 & 551   & 69.6  & 10.79\\   \hline 
		600 & 1.18   & 29.31 & 556   & 70.0  & 11.47\\   \hline
		1200& 1.21 & 30.13 & 558   & 70.3  & 11.84\\   \hline
		2200& 1.20  & 30.00 & 557   & 70.3  & 11.76\\  			
		\hline
		\hline	
	\end{tabular} 
\end{center}
\end {table}

\subsection{Optimal solar cell: Homogeneous bandgap \& periodically corrugated backreflector}\label{sec:opto_elechomo_PCBR}

Next, we carried out the optoelectronic optimization of  solar cells  with a homogeneous CZTSSe layer ($A=0$), 
as in Sec.~\ref{sec:opto_elechomo_FBR}, but with a periodically corrugated backreflector. 
The parameter space for  optimizing $\eta$  was set up as: $\Ls\in[100,2200]$~nm, $\ego\in[0.91, 1.49]$~eV for the optical part,  $\Lg\in[1, 550]$~nm, $\zeta\in(0,1)$, and $\Lx\in[100,1000]$~nm.

The values of  $\Jsc$, $\Voc$, $FF$, and $\eta$ predicted by the coupled optoelectronic   
model are presented in Table~\ref{tab-homo-PCBR} for seven representative values of $\Ls$. The values of  $\ego$,  $\Lg$, 
$\zeta$ and $\Lx$  for the optimal designs are also provided in the same table.

On comparing Tables~\ref{tab-homo-FBR} and \ref{tab-homo-PCBR}, we found that periodic corrugation of
the Mo backreflector slightly improves $\eta$ for $\Ls \in[100, 600]$~nm.  For example,
relative to the planar backreflector, the efficiency increases from $10.38\%$ to $10.72\%$
when $\Ls=300$~nm, the other parameters being $\Lg=100$~nm, $\zeta=0.5$,
$\Lx =500$~nm,  and $\ego=1.20$~eV. No improvement in efficiency was found
for $\Ls>600$~nm by the use of a periodically corrugated backreflector. The optimal bandgap of  CZTSSe   remains the same as with the planar backreflector  in Sec.~\ref{sec:opto_elechomo_FBR}; also, the optimal corrugation parameters lie
in narrow ranges: $\Lg\in[99,105]$~nm, $\zeta\in[0.5,51]$, and $\Lx\in[500,510]$~nm.

\begin {table}[h]
\caption {\label{tab-homo-PCBR} 
Predicted parameters of the optimal CZTSSe solar cell with a specified value of $\Ls\in[100,\red{2200}]$~nm, 
		when the CZTSSe layer is homogeneous ($A=0$) and the Mo backreflector is periodically corrugated. The values of $\ego$ provided pertain to the optical part of the model.  } 
\begin{center} 
		\begin{tabular}{ |p{1.2cm}| p{1.2cm}|p{1.2cm}|p{1.2cm}|p{1.2cm}|p{1.8cm}|p{1.2cm}|p{1.2cm}|p{1.2cm}|}
			\hline
			\hline
			$\Ls$&   $\ego$ &$\Lg$& $\zeta$&$\Lx$&	$\Jsc$ & $\Voc$&$FF$&$\eta$  \\
			(nm) & (eV)&(nm)&&(nm)& (mA~cm$^{-2}$) & (mV)& (\%)& (\%) \\
			\hline
			100 &   1.21   &100&0.50&500& 19.99 & 506   & 75.2  & 7.62	\\  \hline
			200 &  1.20   &105&0.51&510& 25.34 & 532   & 72.3  & 9.75\\ \hline
			300 &   1.20   &100&0.50&500& 27.87 & 546   & 70.4  & 10.72\\ \hline
			400  & 1.19   &103&0.51&502& 28.56 & 547   & 69.7  & 10.91\\   \hline 
			600 &    1.18   &99 &0.50&508& 29.43 & 556   & 70.2  & 11.50\\   \hline
			1200&    1.21   &101&0.51&500& 30.13 & 558   & 70.3  & 11.84\\   \hline
			2200&  1.20   &100&0.50&500& 30.00 & 557   & 70.3  & 11.76\\  		
			\hline
			\hline	
	\end{tabular} 
\end{center}
\end {table}

\subsection{Optimal solar cell: Linearly graded bandgap and planar/periodically corrugated backreflector} \label{sec:Linearly_nonhomo}

Next, we considered the maximization of $\eta$ 
when the bandgap of the CZTSSe  layer is linearly  graded, according to either Eq.~\r{Eqn:Linear-bandgap}
or Eq.~\r{Eqn:Linear-bandgap1}, and the  backreflector is either planar or periodically corrugated. 

\subsubsection{Backward  grading}

Equation~\r{Eqn:Linear-bandgap} is used for backward grading,  i.e., the bandgap near the front contact
is larger than the bandgap near the back contact for $A>0$. 
Optoelectronic optimization yielded $A=0$, i.e., a homogeneous bandgap, whether
the backreflector is planar or periodically corrugated. Therefore, the optimized
results provided in Secs.~\ref{sec:opto_elechomo_FBR} and 
\ref{sec:opto_elechomo_PCBR} also apply for backward bandgap grading of the
CZTSSe layer.

\subsubsection{Forward  grading} \label{sec:Forward-grading}

On the other hand, when  Eq.~\r{Eqn:Linear-bandgap1} is used,  the bandgap near the front contact is smaller
than the bandgap near the back contact for $A>0$. The parameter space  used for
optimizing $\eta$ is: $\Ls\in[100,2200]$~nm, $\ego\in[0.91, 1.49]$~eV, $\egmax\in[0.91, 1.49]$~eV, $A\in[0,1]$, $\Lg\in[1, 550]$~nm, $\zeta\in(0,1)$, and $\Lx\in[100,1000]$~nm with the condition that $\egmax\ge\ego$.
The values of  $\Jsc$, $\Voc$, $FF$, and $\eta$ predicted by the coupled optoelectronic   
model are presented in Table~\ref{tab-Lin-PCBR} for seven representative values of $\Ls$.
The values of $\ego$, $\egmax$, $A$,  $\Lg$, $\zeta$ and $\Lx$  for the optimal designs are also provided in the same table. The corresponding data for optimal solar cells with a planar backreflector ($\Lg=0$) are provided for comparison in Table~\ref{tab-Lin-FBR}.

Just as in Sec.~\ref{sec:opto_elechomo_PCBR}, on comparing Tables~\ref{tab-Lin-PCBR} and \ref{tab-Lin-FBR}, 
we found that periodic corrugation of the Mo backreflector 
slightly improves $\eta$ for $\Ls \lesssim 600$~nm.   Thus, for 
$\Ls=200$~nm, the maximum efficiency predicted is $11.04$\% with a planar backreflector and 
$11.69$\% with a periodically corrugated backreflector. Whether the backreflector is planar or periodically corrugated,
the optimal parameters for forward grading  are: $\ego=0.92$~eV, $\egmax=1.49$~eV, and $A\approx1$. 
The optimal parameters for the periodically corrugated backreflector for $\Ls=200$~nm are: 
$\Lg=100$~nm, $\zeta=0.50$, and $\Lx=500$~nm. No improvement in efficiency was found
for $\Ls>600$~nm by the use of a periodically corrugated backreflector. 

The highest efficiency predicted 
in Tables~\ref{tab-Lin-PCBR} and \ref{tab-Lin-FBR} is $17.07$\%, which arises when $\Ls=2200$~nm,
$\ego=0.91$~eV, $\egmax=1.49$~eV, and $A=0.99$ for both planar ($\Lg=0$)
and periodically corrugated backreflectors.
The values of $\Jsc$, $\Voc$, and $FF$ corresponding to this 
optimal design are $36.72$~mA~cm$^{-2}$, $628$~mV, and $74.0\%$, respectively.
Relative to the optimal homogeneous CZTSSe layer (Sec.~\ref{sec:opto_elechomo_FBR}),
the maximum efficiency increases from $11.84\%$ to $17.07\%$  
(a relative increase of $44.1\%$) with forward grading of the CZTSSe layer;
concurrently, $\Jsc$, $\Voc$, as well as $FF$ are also enhanced.

The optimal values of $\ego\in[0.91,0.93]$~eV and $A\in[0.99,1.0]$ in Tables~\ref{tab-Lin-PCBR} and \ref{tab-Lin-FBR}, and the optimal values of $\egmax$  are independent of $\Ls$, whether the backreflector is planar  or periodically corrugated. Also, the optimal corrugation parameters are very weakly dependent on $\Ls$: $\Lg=100$~nm, $\zeta\in[0.5,51]$, and $\Lx\in[500,550]$~nm.
\begin {table}[h]
\caption {\label{tab-Lin-PCBR} 
	{	Predicted parameters of the optimal CZTSSe solar cell with a specified value of $\Ls\in[100,\red{2200}]$~nm, 
		when the CZTSSe layer is linearly graded ($A\neq0$) according to Eq.~(\ref{Eqn:Linear-bandgap1}) and the Mo backreflector
		is periodically corrugated. The values of $\ego$ and $\egmax$
		provided pertain to the optical part of the model. }} 
\begin{center} {
		\begin{tabular}{ |p{1.0cm}|p{1.0cm}|p{1.0cm}|p{1.0cm}|p{1.0cm}|p{1.0cm}|p{1.0cm}|p{1.8cm}|p{1.0cm}|p{1.0cm}|p{1.0cm}|}
			\hline
			\hline
			$\Ls$&$\ego$&$\egmax$& $A$ &$\Lg$&$\zeta$&$\Lx$&	$\Jsc$ & $\Voc$&$FF$&$\eta$  \\
			(nm) & (eV)&(eV)&&(nm)&&(nm)& (mA~cm$^{-2}$) & (mV)& (\%)& (\%) \\
			\hline
			100 & 0.92 &1.49  &0.99 &100 &0.50 &510 & 20.24 & 544   & 76.0  & 8.44	\\  \hline
			200 & 0.92 &1.49  &1.00 &100 &0.50 &500 & 27.42 & 572   & 74.5  & 11.69 \\ \hline
			300 & 0.91 &1.49  &0.99 &100 &0.50 &510& 29.88  & 592   & 74.0  & 13.01\\ \hline
			400 & 0.92 &1.49  &0.99 &100 &0.51 &550 & 31.39 & 603   & 73.0  & 13.91\\   \hline 
			600 & 0.91 &1.49  &1.00 &100 &0.50 &502 & 32.98 & 612   & 73.6  & 14.87\\   \hline
			1200& 0.93 &1.49  &0.99 &100 &0.51 &500 & 35.02 & 617   & 73.5  & 15.90\\   \hline
			2200& 0.91 &1.49  &0.99 &100 &0.51 &500 & 36.72 & 628   & 74.0  & 17.07\\  			
			\hline
			\hline	
	\end{tabular} }
\end{center}
\end {table}

\begin {table}[h]
\caption {\label{tab-Lin-FBR} 
	Predicted parameters of the optimal CZTSSe solar cell with a specified value of $\Ls\in[100,\red{2200}]$~nm, 
		when the CZTSSe layer is linearly graded ($A\neq0$) according to Eq.~(\ref{Eqn:Linear-bandgap1}) and the Mo backreflector 
		is planar ($\Lg=0$). The values of $\ego$ and $\egmax$
		provided pertain to the optical part of the model.
		} 
\begin{center} 
		\begin{tabular}{ |p{1.0cm}|p{1.0cm}|p{1.0cm}|p{1.0cm}|p{1.8cm}|p{1.0cm}|p{1.0cm}|p{1.0cm}|}
			\hline
			\hline
			$\Ls$&$\ego$&$\egmax$& $A$ &	$\Jsc$ & $\Voc$&$FF$&$\eta$  \\
			(nm) & (eV)&(eV)&& (mA~cm$^{-2}$) & (mV)& (\%)& (\%) \\
			\hline
			100 & 0.91 &1.49  &0.99 & 19.34 & 550   & 76.8  & 8.18	\\  \hline
			200 & 0.92 &1.49  &0.99 & 26.18 & 568   & 74.2  & 11.04\\ \hline
			300 & 0.91 &1.49  &0.99 & 30.07 & 590   & 73.2  & 13.00\\ \hline
			400 & 0.91 &1.49  &0.99 & 31.16 & 601   & 73.4  & 13.75\\   \hline 
			600 & 0.92 &1.49  &0.99 & 33.17 & 610   & 73.6  & 14.92\\   \hline
			1200& 0.93 &1.49  &0.99 & 35.02 & 617   & 73.5  & 15.90\\   \hline
			2200& 0.91 &1.49  &0.99 & 36.72 & 628   & 74.0  & 17.07\\  			
			\hline
			\hline	
	\end{tabular} 
\end{center}
\end {table}

\begin{figure}[h]
		\centering   
		\includegraphics[width=0.9\columnwidth]{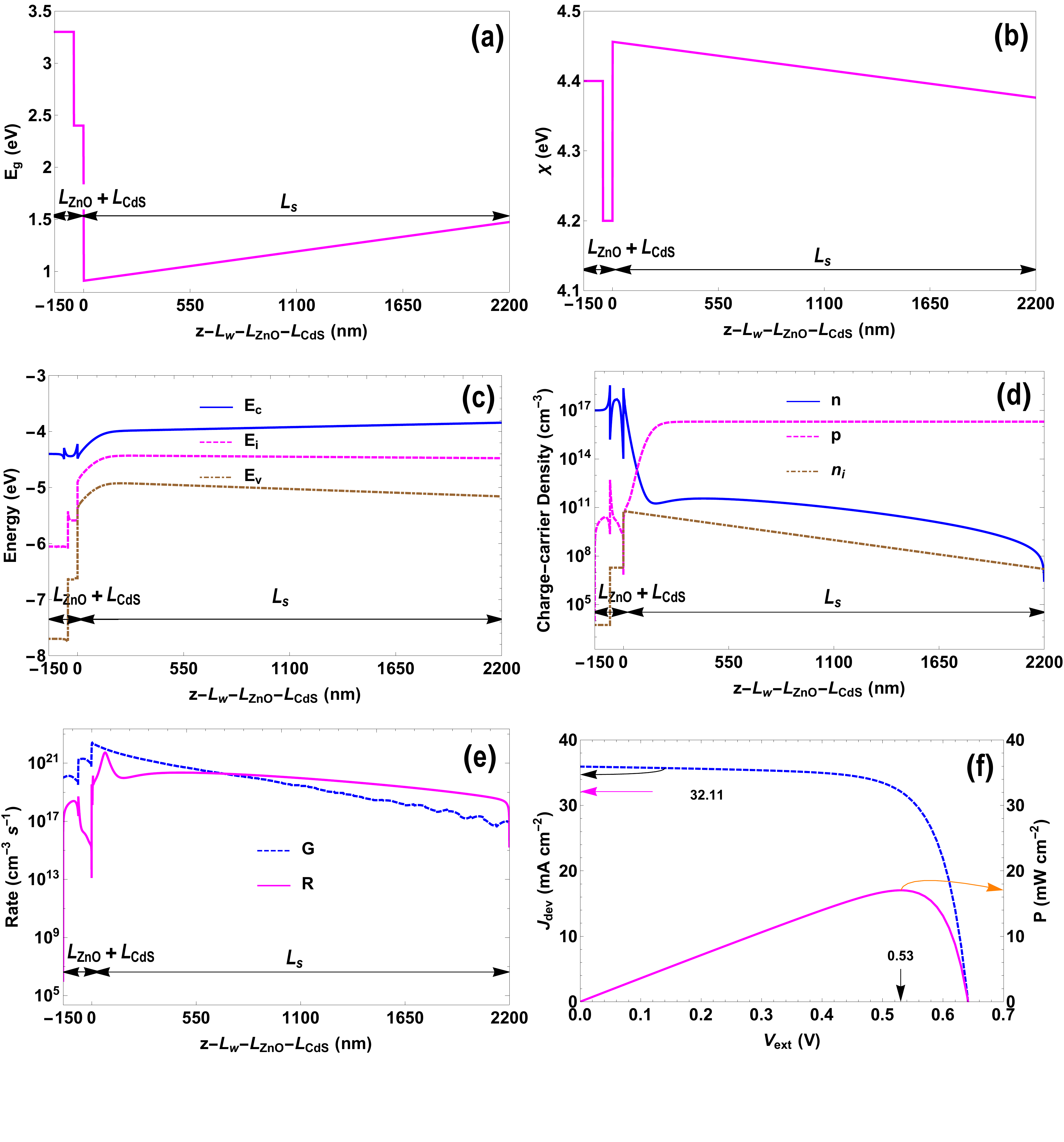} 
		\caption{Spatial profiles of (a) $\eg(z)$; (b) $\chi(z)$; (c) $\Ec(z)$, $\Ev(z)$, and $\Ei(z)$;
		(d) $n(z)$, $p(z)$, 
			and $\ni(z)$; and (e) $G(z)$ and $\Rnpz$ 
			in the semiconductor region of the optimal solar cell with the $2200$-nm-thick 
			CZTSSe layer with forward-graded bandgap.
			(f)   $\Jdev$-$\Vext$ and $P$-$\Vext$ curves   of this
			solar cell. The numerical values of $\Jdev$ and $\Vext$ for maximum $P$ 
			are also identified. {The spatial profiles are the same whether the Mo backreflector
			is periodically corrugated (Table~\ref{tab-Lin-PCBR}) or planar (Table~\ref{tab-Lin-FBR}).}}
			\label{figure:EgXiEcEvEinpniGRJV-linNonH} 
	\end{figure}

{No difference could be discerned in the semiconductor regions of the  forward-graded
solar cells with the highest efficiency in Tables~\ref{tab-Lin-PCBR} and \ref{tab-Lin-FBR},
the CZTSSe absorber layer being $2200$-nm thick whether the backreflector
is planar or periodically corrugated.
Spatial profiles of $\eg(z)$ and $\chi(z)$ are} provided in Fig.~\ref{figure:EgXiEcEvEinpniGRJV-linNonH}(a,b), whereas Fig.~\ref{figure:EgXiEcEvEinpniGRJV-linNonH}(c) presents the spatial profiles of  
$\Ec(z)$, $\Ev(z)$, and $\Ei(z)$. The spatial variations of $\Ec$ and $\Ei$ are quasilinear, quite similar to that
of $\eg$. Figure~\ref{figure:EgXiEcEvEinpniGRJV-linNonH}(d) presents the spatial profiles of $n(z)$, $p(z)$, and $\ni(z)$. We note that $\ni$ varies linearly with $z$ such that it is small where $\eg$ is large and \textit{vice versa}. 

Spatial profiles of $G(z)$ and $\Rnpz$ are provided in Fig.~\ref{figure:EgXiEcEvEinpniGRJV-linNonH}(e).
The  generation rate is higher near the front face and lower near the rear face of the CZTSSe layer, 
which is in accord with the understanding~\cite{Fonash,Repins2016} {that more charge carriers are generated in regions} where  $\eg$ is lower and \textit{vice versa}; {less energy is required to excite a charge carrier from the valence band to the conduction band when $\eg$ is lower.}
The $\Jdev$-$\Vext$ curve of the solar cell is shown in 
Fig.~\ref{figure:EgXiEcEvEinpniGRJV-linNonH}(f). 
From this figure, $\Jdev=32.11$~mA~cm$^{-2}$, $\Vext=0.53$~V,  and 
$FF=74.0\%$ for best performance.	
	
\subsection{Optimal solar cell: Sinusoidally graded bandgap \& planar/periodically corrugated backreflector}\label{sec:optoelec_nonhomo_PCBR}

Finally, we considered the maximization of  $\eta$ for solar cells
with a sinusoidally graded CZTSSe layer according to Eq.~\r{Eqn:Sin-bandgap} and a periodically 
corrugated backreflector. The parameter space   used for  optimizing $\eta$ is: $\Ls\in[100,2200]$~nm, $\ego\in[0.91, 1.49]$~eV, $A\in[0,1]$,  $\alpha\in[0, 8]$, $K\in[0, 8]$, $\psi\in[0,1]$, $\Lg\in[1, 550]$~nm, $\zeta\in(0,1)$, and $\Lx\in[100, 1000]$~nm.
{The values of  $\Jsc$, $\Voc$, $FF$, and $\eta$ predicted by the coupled optoelectronic model are presented 
in Table~\ref{tab-Sin-PCBR} for eight representative values of $\Ls$.
The values of $\ego$, A, K, $\alpha$, $\psi$, $\Lg$, $\zeta$ and $\Lx$  for the optimal designs are also provided 
in the same table. For comparison, the corresponding data for optimal solar cells with a planar backreflector   ($\Lg=0$) are provided in Table~\ref{tab-Sin-FBR}. }

{Just as in Secs.~\ref{sec:opto_elechomo_PCBR} and \ref{sec:Forward-grading},  on comparing Tables~\ref{tab-Sin-PCBR} and \ref{tab-Sin-FBR}, }
we found that periodic corrugation of the Mo backreflector 
slightly improves $\eta$ for $\Ls \lesssim 600$~nm.
For $\Ls=200$~nm, the optimal efficiency predicted is $17.48\%$ with 
a planar backreflector {(Table~\ref{tab-Sin-FBR})} and $17.83\%$ with a periodically corrugated backreflector {(Table~\ref{tab-Sin-PCBR})}. The optimal bandgap parameters for either backreflector are: {$\ego=0.92$~eV,} $A=0.99$, $\alpha=6$, $K=3$, and $\psi=0.75$. The geometric parameters of the optimal periodically corrugated backreflector are: $\Lg=100$~nm, $\zeta=0.51$, and $\Lx=510$~nm. 
For $\Ls=2200$~nm, the optimal efficiency predicted is $19.56\%$, regardless of the geometry
of the backreflector,  the optimal bandgap  parameters being: $\ego=0.92$~eV, $A=0.98$,  $\alpha=6$, $K=2$, and $\psi=0.75$. Indeed,
the effect of periodic corrugation remains the same as in the cases of the homogeneous bandgap (Sec.~\ref{sec:opto_elechomo_PCBR}) and the  linearly graded
bandgap (Sec.~\ref{sec:Forward-grading}): very small improvement for thin CZTSSe layers and no improvement beyond $\Ls\simeq600$~nm. 

{The optimal designs in Table~\ref{tab-Sin-PCBR} have $\Lg\in[100,110]$~nm, $\zeta\in[0.5,0.51]$ and $\Lx\in[500,510]$~nm. 
The values of $\ego\in[0.91,0.92]$~eV, $A\in[0.98,0.99]$, $\alpha=6$, $\psi=0.75$, and $K\in\left\{2,3\right\}$ for both 
planar and periodically corrugated backreflectors. }

\begin {table}[h]
\caption {\label{tab-Sin-PCBR} 
{	Predicted parameters of the optimal CZTSSe solar cell with a specified value of $\Ls\in[100,\red{2200}]$~nm, 
		when the CZTSSe layer is sinusoidally graded ($A\neq0$) according to Eq.~(\ref{Eqn:Sin-bandgap}) and the Mo backreflector is periodically corrugated. The values of $\ego$  
		provided pertain to the optical part of the model.}} 
\begin{center} {
		\begin{tabular}{ |p{1.0cm}|p{1.0cm}|p{0.6cm}|p{0.6cm}|p{0.6cm}|p{0.6cm}|p{0.6cm}|p{0.6cm}|p{0.6cm}|p{1.0cm}|p{0.8cm}|p{0.8cm}|p{0.8cm}|}
			\hline
			\hline
			$\Ls$&$\ego$& $A$ & $K$ &$\alpha$&$\psi$&$\Lg$&$\zeta$&$\Lx$&	$\Jsc$ & $\Voc$&$FF$&$\eta$  \\
			(nm) & (eV)&&&&&(nm)&&(nm)& (mA & (mV)& (\%)& (\%) \\
			 & &&&&&&&& cm$^{-2}$) & & &  \\
			\hline
			100 & 0.92   &0.98&3&6&0.75&100 &0.50 &500 & 25.72 & 701   & 78.7  & 14.22	\\  \hline
			200 & 0.92   &0.99&3&6&0.75&100 &0.51 &510 & 32.99 & 716   & 77.5  & 17.83\\ \hline
			300 & 0.92   &0.98&2&6&0.75&100 &0.51 &510 & 35.15 & 745   & 74.7  & 19.58\\ \hline
			400 & 0.92   &0.98&2&6&0.75&100 &0.51 &510 & 36.32 & 762   & 74.4  & 20.62\\   \hline 
			600 & 0.92   &0.98&2&6&0.75&100 &0.50 &500 & 37.23 & 771   & 74.8  & 21.47\\   \hline
			870 & 0.92   &0.98&2&6&0.75&100 &0.50 &500 & 37.39 & 772   & 75.2  & 21.74\\   \hline
			1200& 0.92   &0.98&2&6&0.75&100 &0.51 &510 & 37.08 & 766   & 74.8  & 21.26\\   \hline
			2200& 0.92   &0.98&2&6&0.75&100 &0.51 &510 & 36.45 & 736   & 72.8  & 19.56\\  			
			\hline
			\hline	
	\end{tabular} }
\end{center}
\end {table}

\begin {table}[h]
\caption {\label{tab-Sin-FBR} 
	{	Predicted parameters of the optimal CZTSSe solar cell with a specified value of $\Ls\in[100,\red{2200}]$~nm, 
		when the CZTSSe layer is sinusoidally graded ($A\neq0$) according to Eq.~(\ref{Eqn:Sin-bandgap}) and the Mo backreflector is planar ($\Lg=0$). The values of $\ego$  
		provided pertain to the optical part of the model.}} 
\begin{center} {
		\begin{tabular}{ |p{1.0cm}|p{1.0cm}|p{1.0cm}|p{1.0cm}|p{1.0cm}|p{1.0cm}|p{1.2cm}|p{1.0cm}|p{1.0cm}|p{1.0cm}|}
			\hline
			\hline
			$\Ls$&$\ego$& $A$ &$K$&$\alpha$&$\psi$&	$\Jsc$ & $\Voc$&$FF$&$\eta$  \\
			(nm) & (eV)&&&&& (mA & (mV)& (\%)& (\%) \\
			& &&&&& cm$^{-2}$) & & &  \\
			\hline
			100 & 0.91   &0.99&3&6&0.75 & 25.65 & 703   & 78.6  & 14.19	\\  \hline
			200 & 0.92   &0.99&3&6&0.75 & 32.40 & 719   & 75.0  & 17.48\\ \hline
			300 & 0.92   &0.98&2&6&0.75 & 33.94 & 744   & 75.0  & 19.01\\ \hline
			400 & 0.92   &0.98&2&6&0.75 & 35.69 & 762   & 75.0  & 20.35\\   \hline 
			600 & 0.92   &0.98&2&6&0.75 & 37.17 & 771   & 74.8  & 21.46\\   \hline
			870 & 0.92   &0.98&2&6&0.75 & 37.39 & 772   & 75.2  & 21.74\\   \hline
			1200& 0.92   &0.98&2&6&0.75 & 37.08 & 766   & 74.8  & 21.26\\   \hline
			2200& 0.92   &0.98&2&6&0.75 & 36.45 & 736   & 72.8  & 19.56\\  			
			\hline
			\hline	
	\end{tabular} }
\end{center}
\end {table}

\begin{figure}[h]
	\centering   
	\includegraphics[width=0.7\columnwidth]{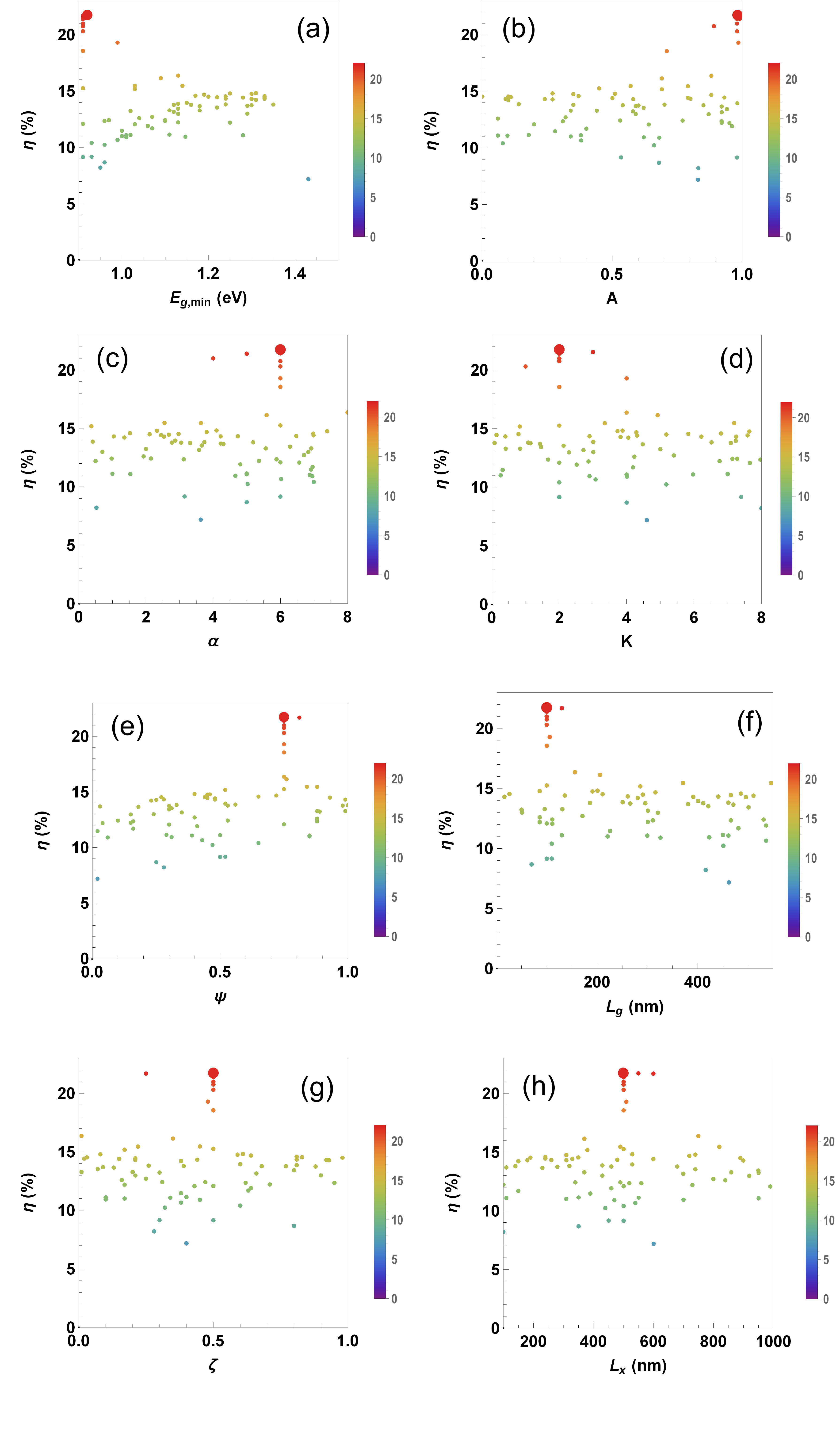}  
\caption{Scatter plots of the optimization results projected onto the plane containing $\eta$ and  (a) $\ego$, (b) $A$, (c) $\alpha$, (d) $K$, (e) $\psi$, (f) $\Lg$, (g) $\zeta$, and (h) $\Lx$, for solar cells with a $870$-nm-thick sinusoidally graded CZTSSe layer and a periodically corrugated backreflector. The large dots highlight the location of the solar cell with the maximum efficiency.} \label{Figure4} 
\end{figure}

The highest efficiency achievable is predicted to be $21.74$\% with a
sinusoidally graded CZTSSe layer of thickness $\Ls=870$~nm, whether the
backreflector is planar {(Table~\ref{tab-Sin-FBR})} or
periodically corrugated {(Table~\ref{tab-Sin-PCBR})}. 
Figure~\ref{Figure4} shows the projections of the    nine-dimensional  space onto the sets
of axes with the efficiency on the vertical axis and each of the optimization parameters on the 
horizontal axis, when $\Ls=870$~nm and the backreflector is periodically corrugated.
The large dots highlight the location of the solar cell with the maximum efficiency. The optimal combination of the values of the parameters
$\ego$, $A$, $\alpha$, $K$,$\psi$, $\Lg$, $\zeta$, and $\Lx$ is recorded in Table~\ref{tab-Sin-PCBR}.

The highest possible efficiency ($21.74\%$) with a sinusoidally graded CZTSSe layer amounts to
a relative increase of $83.6\%$ over the optimal efficiency of $11.84\%$ with a homogeneous CZTSSe layer of 
thickness $\Ls=1200$~nm   (Secs.~\ref{sec:opto_elechomo_FBR} and \ref{sec:opto_elechomo_PCBR}). 
Along with the increase in efficiency, $\Jsc$ increases from $30.13$ mA cm$^{-2}$ to $37.39$ mA cm$^{-2}$ 
(a relative increase of $24.0\%$), $\Voc$ from $558$~mV to
$772$~mV (a relative increase of $38.3\%$), and $FF$ from $70.3\%$ to $75.2\%$ (a relative increase of $6.9\%$). 

The highest possible efficiency ($21.74\%$) with a sinusoidally graded CZTSSe layer
is $27.3\%$ higher than the highest possible efficiency ($17.07\%$) with a linearly graded
 CZTSSe layer 
(Sec.~\ref{sec:Forward-grading}).
The short-circuit current density for sinusoidal  grading is somewhat higher as well,
but the open-circuit voltage is enhanced considerably from  $628$~mV to $772$~mV. Let us note,
however, that the optimal sinusoidally graded CZTSSe layer is only $870$-nm thick,
but its optimal linearly graded counterpart is $2200$-nm thick. 
Indeed, the sinusoidally graded bandgap is more efficient than the homogeneous and linearly graded
bandgaps for all considered thicknesses of the CZTSSe layer.

\begin{figure}[h]
			\centering   
			\includegraphics[width=0.7\columnwidth]{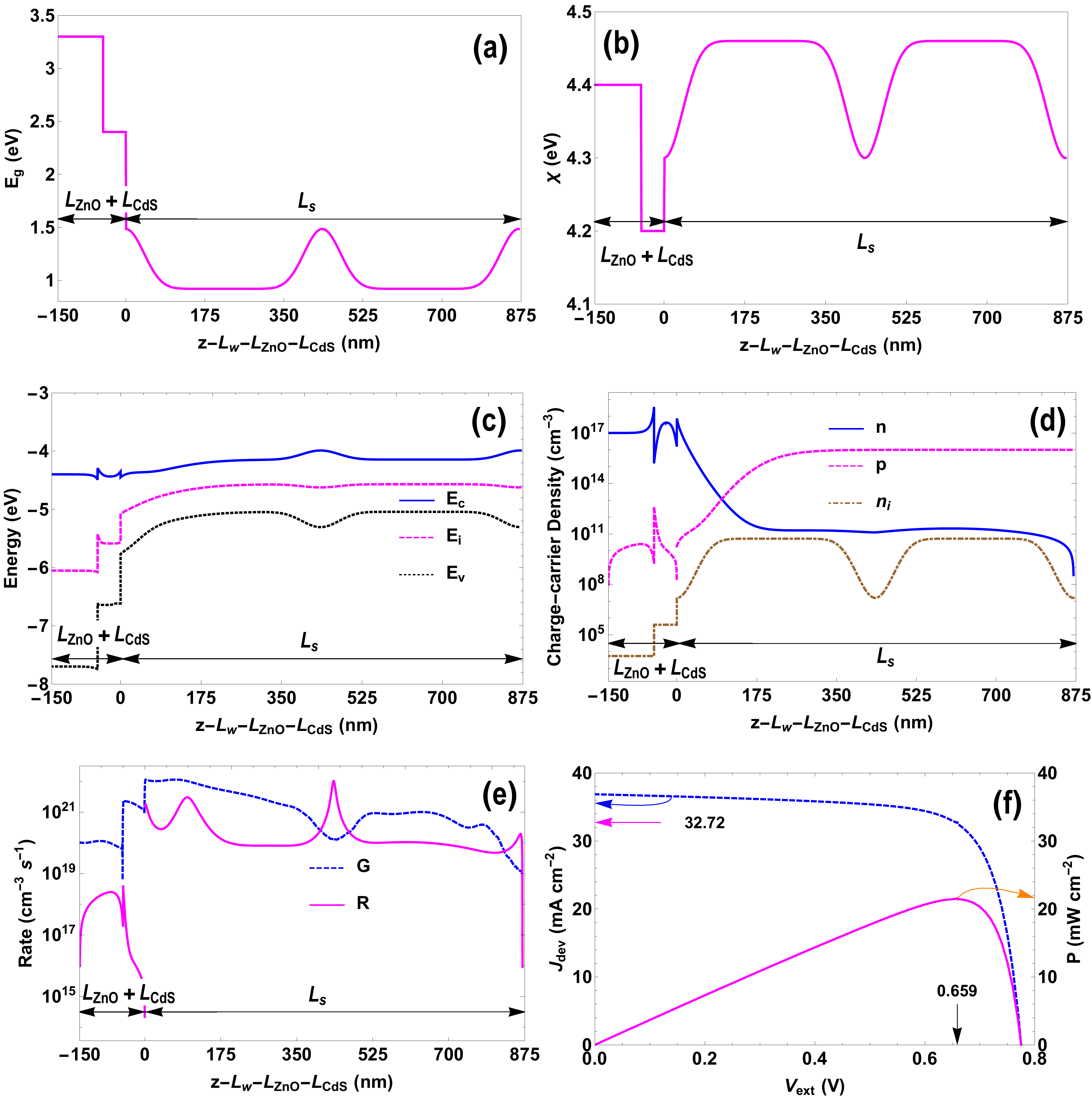} 
			\caption{Spatial profiles of (a) $\eg(z)$; (b) $\chi(z)$; (c) $\Ec(z)$, $\Ev(z)$, and $\Ei(z)$; (d) $n(z)$, $p(z)$, and $\ni(z)$; and (e) $G(z)$ and $\Rnpz$ in the semiconductor region of the optimal solar cell with the $870$-nm-thick CZTSSe layer with sinusoidally graded bandgap. (f)   $\Jdev$-$\Vext$ and $P$-$\Vext$ curves of this solar cell. The numerical values of $\Jdev$ and $\Vext$ for maximum $P$ are also identified. The spatial profiles are the same whether the Mo backreflector is periodically corrugated (Table~\ref{tab-Sin-PCBR}) or planar (Table~\ref{tab-Sin-FBR}).}
				\label{figure:EgXiEcEvEinpniGRJV-SinNonH} 
		\end{figure}

The variations of  $\eg$ and $\chi$  with $z$ in the  semiconductor region of the  solar cell with the optimal sinusoidally graded $870$-nm-thick CZTSSe layer  are provided in Fig.~\ref{figure:EgXiEcEvEinpniGRJV-SinNonH}(a,b). With $\ego=0.92$~eV and $A=0.98$, $\eg(z)\in[0.92,1.486]$~eV. The magnitude of $\eg(z)$ is large near both faces of the CZTSSe layer, which elevates $\Voc$~\cite{Yang2016}. {Furthermore,  bandgap grading in the proximity of the
rear face of the CZTSSe layer keeps the minority carriers away from that face (where
recombination would be highly favored in the absence of the $\Al2O3$ layer~\cite{FLiu2017})
to reduce recombination~\cite{Dullweber2001}  and improve the carrier collection due to the drift field provided by the bandgap grading~\cite{Hutchby1975}.}
The regions in which $\eg$ is  small are of substantial thickness, and it is those very regions that are
responsible for increasing the electron-hole-pair generation rate~\cite{Fonash,Repins2016}, {because less energy is required to excite an electron-hole pair across a narrower bandgap.}
Thus, this  bandgap profile is ideal for the enhancement of $\Voc$ while maintaining a large $\Jsc$.

Figure~\ref{figure:EgXiEcEvEinpniGRJV-SinNonH}(c) shows 
the variations of $\Ec$, $\Ev$, and $\Ei$ with respect to $z$.
The spatial profiles of $\Ec$ and $\Ei$ are  similar to that of $\eg$.
Figure~\ref{figure:EgXiEcEvEinpniGRJV-SinNonH}(d) shows the spatial variations
of   $n$, $p$, and $\ni$  under equilibrium; specifically, $\ni$
varies  such that
it is large  where $\eg$ is small and \textit{vice versa}.	
The spatial profiles of   $G$ and 
$R$ are shown in Fig.~\ref{figure:EgXiEcEvEinpniGRJV-SinNonH}(e). Specifically,
 $G$ is higher in regions with
lower $\eg$ and \textit{vice versa}, as discussed for Fig.~\ref{figure:EgXiEcEvEinpniGRJV-SinNonH}(a).
The higher recombination rate in the  $60$-nm-thick middle region is due to higher defect/trap density caused by higher sulfur content. 
The $\Jdev$-$\Vext$ characteristics are shown in 
Fig.~\ref{figure:EgXiEcEvEinpniGRJV-SinNonH}(f).  Our optoelectronic
model predicts $\Jdev=32.72$~mA~cm$^{-2}$, $\Vext=0.659$~V, and 
$FF=75.2$\% for best performance.

\section{Concluding remarks}\label{sec:conc}
We implemented a coupled optoelectronic model  along with the differential evolution algorithm
to assess the efficacy of grading the bandgap of the
CZTSSe layer
for enhancing the power conversion efficiency  of thin-film  CZTSSe solar cells. Both linearly
and sinusoidally graded bandgaps were examined, with the Mo backreflector
in the solar cell being either planar or periodically corrugated. 
			 
An $870$-nm-thick  sinusoidally graded CZTSSe layer
accompanied by a periodically corrugated  backreflector delivers
 a $21.74$\% efficiency, $37.39$~mA~cm$^{-2}$ short-circuit current density, $772$~mV open-circuit voltage, 
and  $75.2$\% fill factor.
Even if  the backreflector is flattened, these quantities do not alter.
In comparison, $\eta=11.84$\%, $\Jsc=31.13$~mA~cm$^{-2}$, $\Voc=558$~mV, and 
$FF=70.3$\%, when the bandgap is homogeneous and the backreflector is planar. 
Efficiency   can also be enhanced by linearly grading the bandgap, 
but the gain is smaller compared to the case of sinusoidal bandgap grading.

The  generation rate is higher in the broad 
small-bandgap regions than elsewhere in the CZTSSe layer,
when the bandgap is sinusoidally graded.  Since the bandgap is high close to both faces of the CZTSSe 
layer, $\Voc$ is high in the
optimal designs~\cite{Yang2016, Gloeckler-Sites2005}. Both of these features are responsible of enhancing $\eta$.

The placement of an ultrathin $\Al2O3$ layer behind the rear face of the CZTSSe layer helps
remove an unwanted Mo(S$_\xi$Se$_{1-\xi}$)$_2$ layer and slightly enhances the efficiency.   
Furthermore, for a thin CZTSSe layer ($\Ls\leq500$~nm), periodically corrugating the backreflector can also 
provide  small gains over a planar backreflector.

Optoelectronic optimization thus indicates that $21.74$\% efficiency can be achieved for CZTSSe 
solar cell with a $870$-nm-thick CZTSSe layer. This efficiency significantly higher compared to $12.6$\% efficiency 
demonstrated with CZTSSe layers that are  more than two times thicker.  Efficiency enhancements of
comparable magnitude---e.g., $22\%$ to $27.7\%$---have
 been predicted by bandgap grading of the CIGS layer in thin-film CIGS solar cells
\cite{Faiz-CIGS-AO} (which, however, use some materials that are not known to be
abundant on Earth).
Thus, bandgap grading can provide a way to realize more efficient thin-film solar cells for
ubiquitous small-scale  harnessing of solar energy. 
		
\appendix 
\section{Relative permittivities of materials in the optical regime}\label{Appendix A}

\begin{figure}[!t]
	\centering
	\includegraphics[width=0.6\columnwidth]{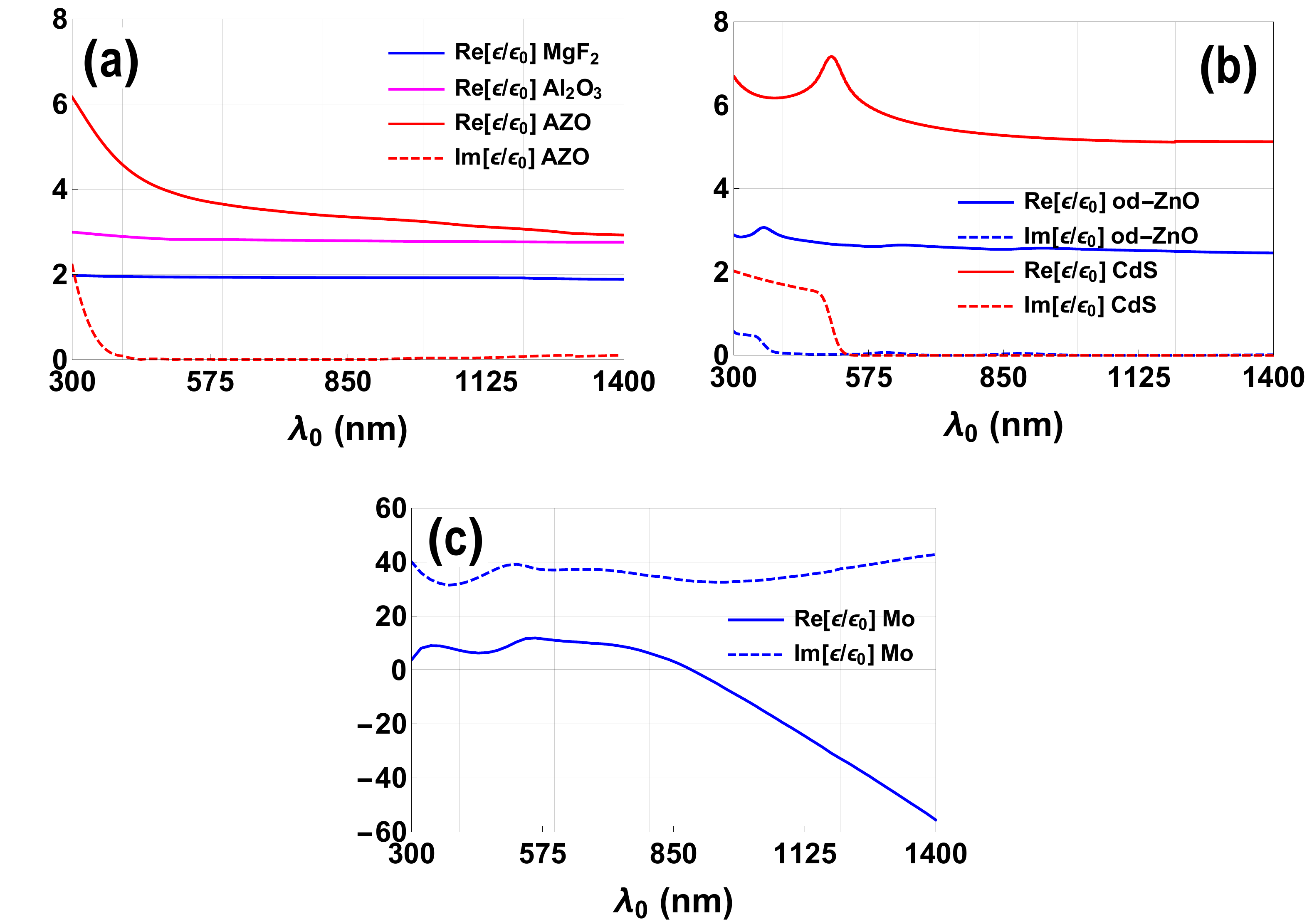}    
	\caption{Spectrums of ${\rm Re}\left[\eps/\epso\right]$ and ${\rm Im}\left[\eps/\epso\right]$
		of (a) $\MgF2$, $\Al2O3$, and AZO, (b)  od-ZnO  and CdS, and (c) Mo.
		${\rm Im}\left[\eps/\epso\right]$ is negligibly small
		for $\MgF2$ and $\Al2O3$.
		 }
		\label{spectra-eps-1}
\end{figure}

\begin{figure}[!t]
	\centering
	\includegraphics[width=0.7\columnwidth]{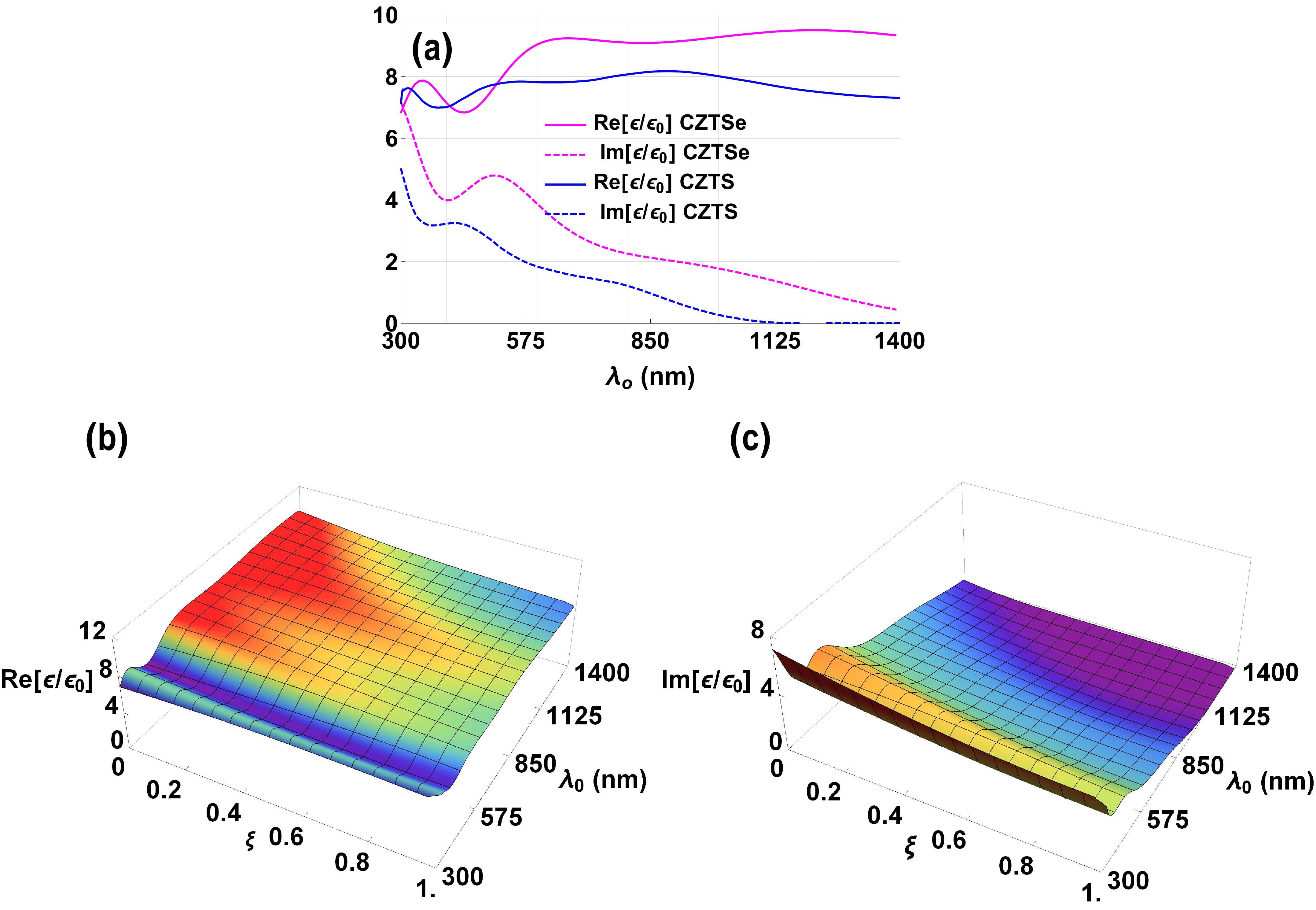}    
	\caption{(a) Spectrums of ${\rm Re}\left[\eps/\epso\right]$ and ${\rm Im}\left[\eps/\epso\right]$
			of CZTS and CZTSe. (b) ${\rm Re}\left[\eps/\epso\right]$ and (c) ${\rm Im}\left[\eps/\epso\right]$  of CZTSSe as functions
		of $\lambdao$ and  $\xi$.}
		\label{CZTSSe-spectra}
\end{figure}

\begin{figure}[!t]
	\centering
	\includegraphics[width=0.7\columnwidth]{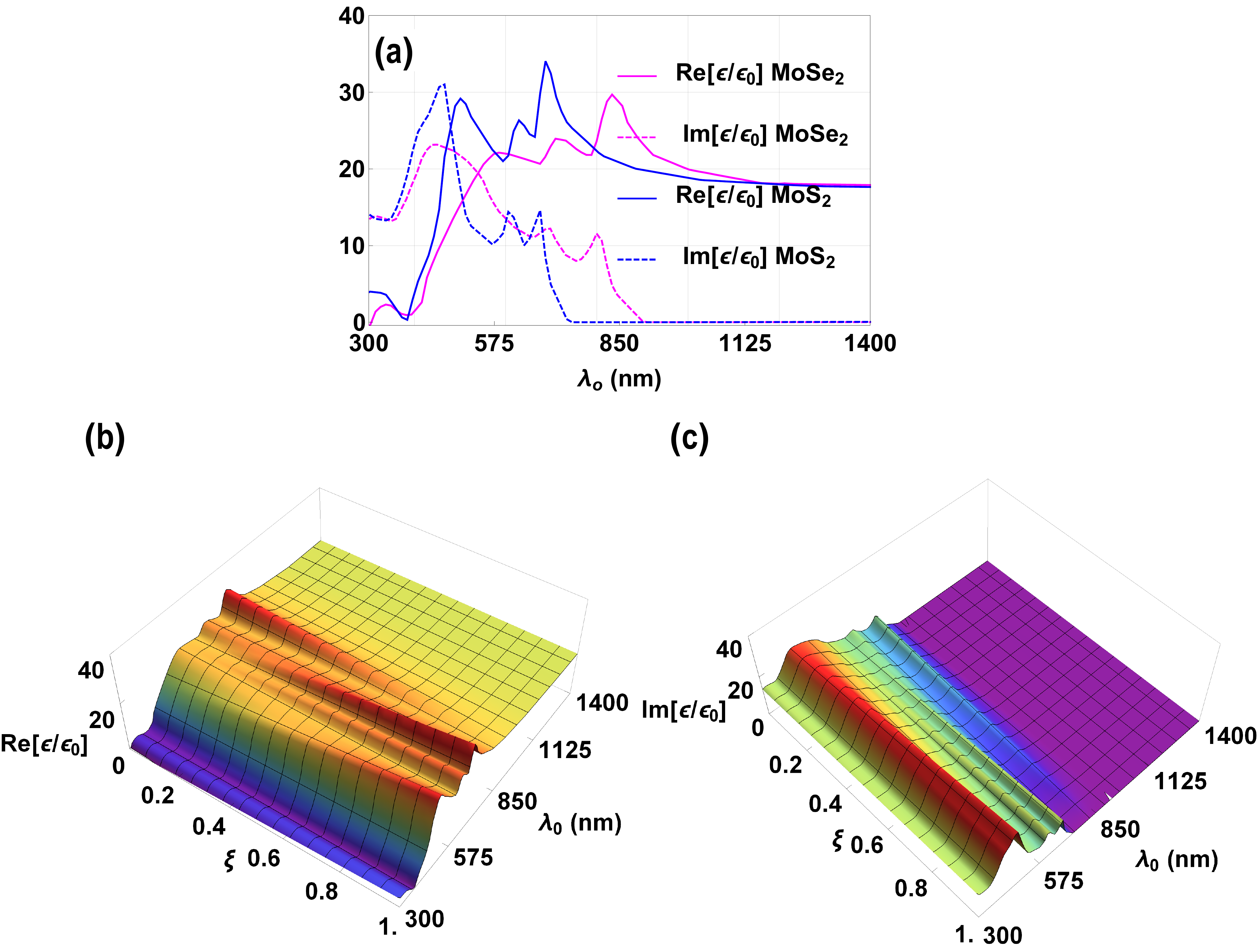}    
	\caption{(a) Spectrums of ${\rm Re}\left[\eps/\epso\right]$ and ${\rm Im}\left[\eps/\epso\right]$
			of MoS$_2$ and MoSe$_2$. (b) ${\rm Re}\left[\eps/\epso\right]$ and (c) ${\rm Im}\left[\eps/\epso\right]$  of Mo(S$_\xi$Se$_{1-\xi}$)$_2$ as functions
		of $\lambdao$ and  $\xi$. }
		\label{MoSSe-spectra}
\end{figure}

Spectra of the real and imaginary parts of the relative permittivity
$\eps(\lambdao)/\epso$ of $\MgF2$~\cite{mgf2}, AZO~\cite{AZO}, od-ZnO~\cite{ZnO}, CdS~\cite{treharne}, 
 Mo \cite{Mo}, and $\Al2O3$~\cite{Al2O3}  in the optical regime are displayed in Fig.~\ref{spectra-eps-1}.
 Spectra of the real and imaginary parts of the relative permittivity
  of  CZTS and CZTSe are available~\cite{Adachi_book2015}. These
were incorporated in an energy-shift model~\cite{Hirate2015,Nakane2016}
to obtain the relative permittivity of CZTSSe  
as a function of $\xi$ (and, therefore, the bandgap $\eg$)
and $\lambdao$ in the optical regime, as shown in Fig.~\ref{CZTSSe-spectra}. 
Spectra of the real and imaginary parts of the relative permittivity
  of  MoS$_2$ and MoSe$_2$ are available
for $\lambdao\leq 1240$~nm~\cite{Beal-Hughes}. These were first linearly extrapolated for
$\lambdao\in(1240,1400]$~nm and
then incorporated in the energy-shift model~\cite{Hirate2015,Nakane2016}
to obtain the relative permittivity of Mo(S$_\xi$Se$_{1-\xi}$)$_2$  
as a function of $\xi$ 
and $\lambdao$, as shown in Fig.~\ref{MoSSe-spectra}. \\

\noindent\textbf{Acknowledgments.} 
The authors thank anonymous reviewers for invaluable suggestions to improve the contents
of this paper. A. Lakhtakia acknowledges the Charles Godfrey Binder 
Endowment at the Pennsylvania State University and the Otto M{\o}nsted Foundation
in Frederiksberg, Denmark for partial support.
The research of  F. Ahmed and A. Lakhtakia was partially supported by  US National 
Science Foundation (NSF) under grant number DMS-1619901. 
The research of  T.H. Anderson and P.B.  Monk was partially supported by  the US
National Science Foundation (NSF) under grant number DMS-1619904.

\end{document}